\documentclass[journal]{vgtc}              

\onlineid{1039}

\vgtccategory{Research}

\usepackage{algorithmic}
\usepackage{algorithm}
\usepackage{array}
\usepackage[caption=false,font=normalsize,labelfont=sf,textfont=sf]{subfig}
\usepackage{url}
\usepackage{verbatim}
\usepackage{ltablex}
\usepackage[table]{xcolor}
\usepackage{caption}

\usepackage{colortbl}
\usepackage{xspace}
\usepackage{soul}
          
\PassOptionsToPackage{warn}{textcomp} 
\usepackage{textcomp}                
\usepackage{multirow}
\usepackage{tabu}
\usepackage{tabularx}
\usepackage{booktabs}
\usepackage{lipsum}
\usepackage{mwe}
\usepackage{mathptmx}

\usepackage{enumitem}
\usepackage{xspace}
\usepackage{mdframed}
\setlist[itemize]{noitemsep, topsep=0pt}
\usepackage{xcolor}

\newcommand{\subhead}[1]{\noindent\textbf{#1}}

\newcommand{\code}[1]{{\scriptsize{#1}}}
\newcommand{\bfcodee}[1]{{\fontsize{8.5pt}{10pt}\selectfont \textbf{#1}}}

\newcommand{\codee}[1]{{\fontsize{8.5pt}{10pt}\selectfont #1}}

\newcommand{\codeee}[1]{{\fontsize{8pt}{10pt}\selectfont #1}}

\definecolor{orange1}{RGB}{242,142,43}
\definecolor{blue1}{RGB}{78,121,167}
\definecolor{yellow1}{RGB}{240,189,39}
\definecolor{pink1}{RGB}{246, 114, 128}
\definecolor{purple1}{RGB}{148,103,189}
\definecolor{green1}{RGB}{89,161,79}
\definecolor{customgray}{RGB}{237, 237, 237}
\definecolor{mag1}{RGB}{181, 115, 157}
\newcommand{\dataset}{\textcolor{orange1}{\bfcodee{dataset}}\xspace}
\newcommand{\model}{\textcolor{blue1}{\bfcodee{model}}\xspace}

\newcommand{\systemOperations}{\textcolor{pink1}{\bfcodee{systemOperations}}\xspace}
\newcommand{\userOperations}{\textcolor{purple1}{\bfcodee{userOperations}}\xspace}

\newcommand{\praxa}{\mbox{\sc Praxa}\xspace}
\newcommand{\type}[1]{#1}

\newcommand{\pslfullname}{\mbox{\sc Praxa Specification Language}\xspace}
\newcommand{\psl}{\mbox{\sc PSL}\xspace}

\newcommand{\dSC}[1]{\textcolor{orange1}{\codee{#1}}}
\newcommand{\mSC}[1]{\textcolor{blue1}{\codee{#1}}}

\newcommand{\sOpSC}[1]{\textcolor{pink1}{\codee{#1}}}
\newcommand{\uOpSC}[1]{\textcolor{purple1}{\codee{#1}}}

\newif\ifnotes
\notestrue

\definecolor{ISColor}{HTML}{FFF2CC}
\definecolor{MPSColor}{HTML}{D9EAD3}
\definecolor{GSColor}{HTML}{CFE2F3}
\definecolor{IColor}{HTML}{F4CCCC}
\definecolor{SColor}{HTML}{D9D2E9}

\definecolor{rowgray}{gray}{0.95} 
\definecolor{rowwhite}{gray}{1}
\title{\praxa: A Grammar for What-If Analysis}

\author{%
  \authororcid{Sneha Gathani}{0000-0002-0706-7166},
  Kevin Li,
  Raghav Thind,
  Sirui Zeng,
  Matthew Xu,
  Peter J. Haas,
  \c{C}a\u{g}atay~Demiralp, and 
  Zhicheng Liu
}

\authorfooter{
    \item
    Sneha Gathani is with University of Maryland, College Park.
    Email: sgathani@umd.edu.
    \item
  	Kevin Li is with University of Maryland, College Park.
  	E-mail: kli36@terpmail.umd.edu.
    \item
  	Raghav Thind is with University of Maryland, College Park.
  	E-mail: rthind@terpmail.umd.edu.
  \item
  	Sirui Zeng is with University of Maryland, College Park.
  	E-mail: szeng124@umd.edu.
  \item Matthew Xu is with University of Maryland, College Park.
  	E-mail: mxu0601@terpmail.umd.edu.
  \item
  	Peter J. Haas is with University of Massachusetts, Amherst.
  	E-mail: phaas@cs.umass.edu.
  \item
  	\c{C}a\u{g}atay~Demiralp is with AWS AI Labs and MIT CSAIL.
  	E-mail: cagatay@csail.mit.edu.
  \item Zhicheng Liu is with University of Maryland, College Park.
  	E-mail: leozcliu@umd.edu.
}

\abstract{%
    What-if analysis is widely used to explore hypothetical scenarios, reason about input–output relationships, and evaluate alternative pathways to desired results. 
However, current approaches to what-if analysis are fragmented: systems implement what-if capabilities under diverse terminologies, and provide different sets of analytic techniques to perform them.
Such fragmentation limits the expressiveness of analytic systems, impedes flexible composition and reuse of analytic workflows, and hinders tighter integration with AI.
We present \praxa, a compositional grammar of what-if analysis derived from recurring structural patterns identified across 141 publications in leading visual analytics and HCI venues.  
\praxa formulates three compositional primitives: (1) data primitive that defines variables over which the analysis is conducted, (2) model primitive that specifies what predictive mechanisms are involved, and (3) interaction operations--pairs of user action and system responses that execute what-if analyses.
To demonstrate the grammar in action, we encode \praxa into a declarative specification language, \psl.
To evaluate \praxa, we first show the expressiveness of the grammar by reconstructing representative what-if workflows from prior work as structured compositions of primitives, and by exposing the limitations of existing literature in their predominant focus on single-step workflows rather than multi-step reasoning.
Second, we demonstrate its composability by revealing that analytic capabilities described under distinct terminologies can be composed using the same underlying grammatical structure with different parameterizations, and that new multi-step workflows can be enabled through composition.
Third, we illustrate \psl as an intermediate representation for translating natural-language what-if queries into executable interactive interfaces, enabling inspection, validation, and more transparent integration with AI-driven what-if workflows.
By unifying diverse what-if analysis approaches as a grammar, \praxa provides a foundation for analyzing, composing, and supporting workflows in next-generation what-if systems.
}


\usepackage{tabu}                      
\usepackage{booktabs}                  
\usepackage{lipsum}                    
\usepackage{mwe}                       
\usepackage{ccicons}                   
\usepackage{mathptmx}                  

\begin{document}



\maketitle
\section{Introduction}
What-if analysis is employed across various visual analytics systems to explore hypothetical scenarios, understand input-output relationships, interpret model behavior under constraints, and evaluate alternative pathways towards achieving desired outcomes.
However, such what-if capabilities are implemented across systems via diverse terminologies, such as scenario modeling~\cite{metz2024interactive,levesque2024pathways,de2023prowis}, sensitivity analysis~\cite{laguna2023explimeable,li2020cnnpruner}, perturbation-based analysis~\cite{liu2018nlize}, what-if analysis~\cite{gathani2022if, sarikaya2019what,lee2022sleepguru,seebacher2021investigating, gathani2021augmenting}, inverse modeling~\cite{orban2018drag,moon2023amortized,raidou2016visual}, and counterfactual analysis~\cite{wu2021principles,kaul2021improving}. 
For example, scenario modeling simulates potential conditions of factors to evaluate different outcomes~\cite{metz2024interactive,waser2014many}. 
Sensitivity analysis assesses the impact of variations in specific factors on outcomes~\cite{borland2024using,evirgen2024text}. 
Perturbation-based analysis deliberately introduces variations to factors and study system behavior~\cite{liu2018nlize}.
Inverse modeling determines the input factors required to produce specific outcomes~\cite{moon2023amortized}, whereas counterfactual analysis identifies minimal changes in input factors necessary to change the model predictions~\cite{shamma2022ev}.
These examples all involve exploring hypothetical scenarios to observe potential outcomes or identifying data and model configurations that achieve desired outcomes, yet they are each described and implemented as isolated techniques.

Such a fragmented landscape makes it difficult to understand the structural similarities and differences between existing approaches.
It is also difficult to extend or compose the supported workflows in a more complex analytic process.
The absence of a shared structural representation also hinders integration with emerging AI-driven analysis pipelines, where systems interpret, generate, and execute what-if workflows from natural language (NL) input.

To address this problem, we present \praxa, a compositional grammar for what-if analysis.
\praxa is derived from a systematic analysis of recurring patterns in the definitions, explanations, and implementations of what-if analysis workflows across 141 publications from leading visual analytics and HCI venues. 
\praxa characterizes what-if workflows through three classes of primitives:
(1) data primitives that specify the variables involved in analysis, 
(2) model primitives that specify the predictive mechanisms, and
(3) interaction operations which include the interplay of user actions and corresponding system responses.
To demonstrate the grammar in action, we encode \praxa into a declarative specification language, \pslfullname (\psl).
The language makes the compositional structure of the grammar explicit, enables primitive compositions as executable what-if workflows, and is independent of domain-specific implementations.

\begin{figure*}[h]
    \centering
    \includegraphics[width=\textwidth]{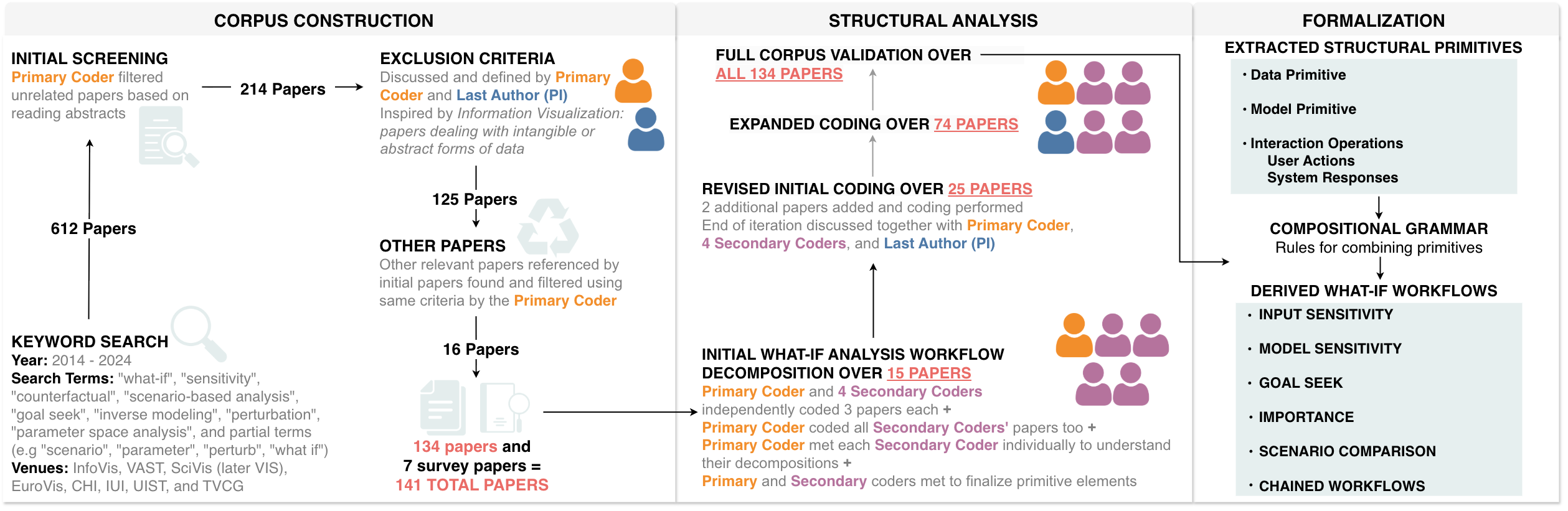}
    \vspace{-5mm}
    \caption{Overview of corpus construction, iterative structural decomposition of what-if workflows, abstraction of recurring components, and formalization into the compositional grammar underlying \praxa.}
    \vspace{-5mm}
    \label{fig:methodology}
\end{figure*}

We then show the utility and evaluate the grammar along three dimensions.
First, we show its expressiveness by mapping what-if analyses observed in prior works into five canonical what-if workflows composed of \praxa's primitives.
This mapping shows that approaches described under diverse terminologies often share the same underlying structure, and reveals that existing systems predominantly support single-step workflows despite users' need for multi-step reasoning.
Second, we demonstrate the grammar's composability by demonstrating how these primitives can be combined to express richer, multi-step workflows. In particular, we surface composition patterns such as chaining and iterative refinement, which are not supported in existing systems but reflect how users naturally conduct continuous what-if analysis.
Finally, we illustrate a practical application of the grammar through \psl, using it as an inspectable intermediate representation for NL what-if interfaces, enabling more transparent and controllable integration of what-if workflows with AI systems.

In summary, we make the following contributions:
\begin{itemize}[left=0.05em]
    \item We conduct a structural analysis of what-if workflows across 141 visual analytics and HCI publications (2014–2024) to derive \praxa, a compositional grammar that formalizes recurring patterns and encodes them in a declarative specification language, \psl.
    
    \item We demonstrate the expressiveness and composability of the grammar by mapping existing what-if approaches into structured compositions of primitives, revealing shared structure across seemingly distinct methods, identifying a bias toward single-step workflows, and exposing underexplored multi-step workflow compositions.
    
    \item We illustrate an application of the grammar by enabling the use of \psl as an intermediate representation for translating NL what-if queries into executable interactive interfaces, enabling structural validation and error detection.
\end{itemize}
\section{Methodology}
\label{sec:methodology}
To derive \praxa, we conducted a systematic structural analysis of what-if workflows across 141 publications from leading visual analytics and HCI venues spanning 2014–2024.
Our objective was to formalize what-if analysis as a compositional grammar rather than a taxonomy of analyses techniques; accordingly, we did not organize prior work by domain, application, or used terminology.
Instead, we treated each what-if feature across prior works as an instance of a workflow and analyzed its components: the data elements over which what-if analysis is conducted, the model elements involved, and the interaction sequence through which user actions and system responses carried out the analysis.
Across works, we identified recurring structural configurations that appeared under diverse terminologies, decomposed them into constituent parts, and
iteratively abstracted until reusable components emerged.
These components formed the primitives of \praxa and motivated the formalization of compositional rules governing how they combine into valid what-if workflows.
Figure~\ref{fig:methodology} summarizes this nine-month process--from corpus construction and workflow decomposition to primitive abstraction and grammar derivation.

\subsection{Corpus Construction}
\label{subsec:corpus_construction}
To capture a breadth of what-if capabilities across prior visual analytics and HCI research, we conducted a keyword-based search for papers published between 2014 and 2024 across InfoVis, VAST, SciVis (later unified as VIS), EuroVis, CHI, IUI, UIST, and TVCG.
Search terms were intentionally expansive to capture a broad range of terms related to what-if analysis, including ``scenario-based analysis'', ``sensitivity'', ``counterfactual'', ``perturbation'', ``what-if'', ``inverse modeling'', ``goal seek'', and related partial terms like ``scenario'' and ``perturb''.
The complete keyword list used is provided in the supplementary materials~\cite{supplementary}.
The search yielded 612 papers.

The first author screened abstracts of all retrieved papers to assess relevance to what-if analysis.
Papers deemed out of scope, like those focused on scientific visualization, purely theoretical discussions, or system optimization without user interaction--were excluded, reducing the corpus to 214 papers. 
The first and last authors then jointly established exclusion criteria for works centered on tangible or physical datasets involving hardware or instrumentation (e.g.,~\cite{wang2018seismo,mukashev2023tacttongue}), further narrowing the corpus to 125 papers.
Finally, backward snowballing over references identified 16 additional relevant works, resulting in 141 papers.
Of these, 7 survey papers were excluded, leaving 134 papers for in-depth analysis.

\subsection{Analysis Workflow Decomposition and Coding Procedure}
\label{subsec:analysis_procedure}
Decomposing the what-if analyses workflows across corpus of papers involved six authors: the first author (primary coder), four additional coders, and the senior author (PI), working iteratively to ensure structural interpretations were consistent across papers and abstracted at a level appropriate for grammar derivation.

\subhead{Initial Workflow Reconstruction.}
First, five coders independently analyzed three papers each (15 papers total), selected to span different what-if terminologies. 
Coders reconstructed each system as one or more complete what-if workflows, documenting: (1) input and output variables, (2) predictive or computational mechanisms connecting them, and (3) the sequence of user actions and corresponding system responses through which the analysis unfolded.
Coders then identified structural primitives in each workflow and described how they were arranged and sequenced to surface recurring workflow patterns.

\subhead{Schema Consolidation.}
To consolidate interpretations, the primary coder re-analyzed the twelve papers reviewed by the additional coders and met individually with each to reconcile workflow reconstructions.
Subsequently, all five coders convened to compare structural descriptions, clarify definitions, and abstract system-specific descriptions into higher-level operations.
Through iterative comparison, three recurring primitives of \praxa were identified across workflows: (1) data primitive, (2) model primitive, and (3) interaction operations involving user actions and system responses. 
This consolidated schema formed the foundation for analyzing the remaining papers.

\subhead{Iterative Coding and Refinement.}
Coding proceeded across three iterations of increasing scale: 25 papers (revisiting the initial 15 and adding 10 new) in the first, 74 total by the second, and the full 134-paper corpus by the third.
After each iteration, all coders and the PI convened to review primitives and its elements together, resolve ambiguities, and refine abstraction levels.
Throughout the iterations, the PI ensured that abstractions remained appropriate for grammar derivation--avoiding overfitting to individual works or systems while being meaningful to distinguish structurally distinct workflow configurations, and composable with one another into larger workflows.

Each coder was responsible for a distinct subset of approximately 27 papers, with the primary coder re-coding additional papers throughout the process as a quality control measure.
Coding each paper required 1.5-2 hours of close reading to reconstruct the workflow structure and map it onto the three primitives.

\begin{figure}[!b]
    \centering
    \includegraphics[width=\columnwidth]{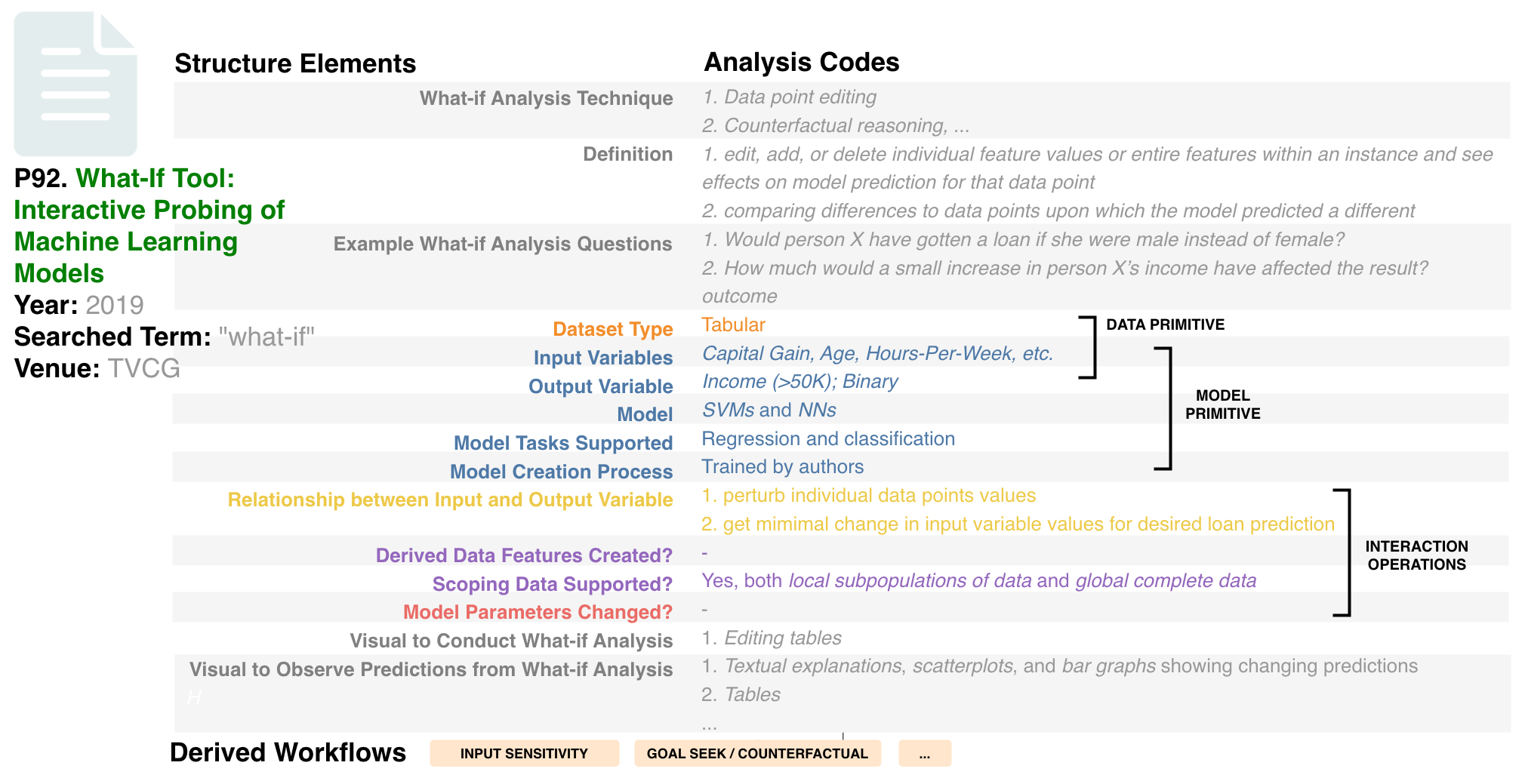}
    \vspace{-5mm}
    \caption{Example of structural decomposition during analysis of Wexler et al.~\cite{wexler2019if}.
    Extracted codes are organized into data, model, and interaction primitives, demonstrating how what-if workflows are decomposed into recurring structural components that underpin \praxa.}
    \vspace{-5mm}
    \label{fig:example}
\end{figure}

\subhead{Synthesis and Grammar Derivation.}
After individual coding was completed, coders also cross-reviewed approximately one-third of each other's papers in pairs to ensure consistency in structural interpretation.
Eventual synthesis sessions then compared workflow decompositions across the full corpus, examining how different systems arrived at structurally distinct
workflow configurations by composing the same underlying primitives in different arrangements.
Figure~\ref{fig:example} illustrates this decomposition using Wexler et al.'s What-If Tool~\cite{wexler2019if}--a system enabling interactive what-if analysis of machine learning models.
The figure shows how the system's two what-if workflows are reconstructed and decomposed into their constituent primitives, revealing that structurally distinct workflows (``data point editing'' and ``counterfactual reasoning'') can be distinguished not by their data and model primitives, which are shared, but by the specific interaction operations that carry out the analysis.

These discussions took place across eight to ten virtual sessions (1.5--2.5 hours each), with three to five additional hour-long sessions
held with all coders to clarify ambiguities and refine primitive assignments.
A key observation to emerge from this synthesis was that many seemingly distinct what-if approaches--described under different terminologies and deployed in different domains--could be reconstructed as different compositions of the same primitives.
This recurring compositionality directly motivated the formalization of \praxa as a grammar: a system of primitives and rules governing how
analytical intents, data and model primitives, and interaction operations combine into valid and feasible what-if workflows.
The finalized codebook is provided in the supplementary materials~\footnotemark[1].
\section{\praxa: A Compositional Grammar of What-If Analysis}
\label{sec:praxa}
\praxa formalizes what-if analysis as a compositional grammar: a system of primitives and rules governing how those primitives combine into valid and feasible what-if workflows.
Through structural analysis of corpus papers, we identified three classes of primitives that recur across what-if workflows: (1) \textit{data primitive}, which define the variables over which what-if workflows are conducted; (2) \textit{model primitive}, which specify the predictive mechanisms involved; and (3) \textit{interaction operations}, which capture the user actions and system responses through which analysis is carried out, which we define underneath.

\subsection{Data Primitives}
\label{subsec:data_primitive}
We found that what-if workflows were applied on various types of datasets: multivariate (68\%; 12\% geo-related/spatial and 2\% time series/temporal), image (10\%), textual data (9\%), audio (1\%), and no datasets (12\%), where the workflow was discussed theoretically or empirical data was not explicitly mentioned or explained.
However, a set of 
\dataset primitives were consistently observed:

\begin{itemize}[left=0.05em]
    \item \dSC{instances}: the fundamental data units (e.g., rows in tabular datasets, individual images in image datasets, or entire documents or prompts in textual datasets).
    \item \dSC{features}: measurable or representational properties associated with each instance, varying by dataset type.
    For example, Wexler et al.~\cite{wexler2019if} perform what-if workflows on tabular dataset with quantitative (e.g., \textit{Capital Gain}), categorical (e.g., \textit{Gender}), and ordinal (e.g., \textit{Capital Gain}) features, Park et al.~\cite{park2021vatun} conduct what-if workflows on pixel intensities and color distributions in images, and Liu et al.~\cite{liu2018nlize} apply workflows on textual datasets having word and sentence embeddings.
    \item \dSC{values}: concrete instantiations of features, such as `Male' or `Female' for \textit{Gender}, `0 to 5000' for \textit{Capital Gain}, RGB channel intensities for images, or token and embedding vectors for text.
\end{itemize}

Together, these elements define the representational space within which what-if workflows are expressed.
\subsection{Model Primitives}
\label{subsec:model_primitive}
The \model primitives represent the predictive or computational mechanism that maps input variables to an output variable.
They define how hypothetical changes introduced by interaction operations are translated into predicted outcomes.
Across the what-if workflows observed in our corpus, different underlying models were used.
However, every workflow instantiated the same structural elements of the \model primitive: the \mSC{inputVariables} drawn from the \dataset, the \mSC{outputVariable} being predicted, the supported \mSC{modelType}, and the associated \mSC{modelParameters}.

\vspace{1.5mm}
\noindent \textbf{3.2.1. Input Variables}

\mSC{inputVariables} are the \dataset \mSC{features} that are considered inputs to the \model for a workflow.
They may correspond to existing dataset features (\mSC{existingFeature}) or be derived from existing features (\mSC{derivedFeature}) through arithmetic or statistical transformations.

\vspace{1.5mm}
\noindent \textbf{3.2.2. Output Variable}

The \mSC{outputVariable} is the dependent variable predicted by the \model.
Like input variables, it may correspond to an existing \dataset feature (\mSC{existingFeature}; e.g., \textit{Income} variable in Wexler et al.~\cite{wexler2019if}), or be derived from existing features (\mSC{derivedFeature}; e.g., \textit{Goodness} score computed through pairwise comparisons between data points).
Across the corpus, what-if workflows were consistently organized around a single output variable.

\vspace{1.5mm}
\noindent \textbf{3.2.3. Model Type}

The type of model (\mSC{modelType}) characterizes the predictive or computational mechanism used in a workflow.
Different workflows across the reviewed corpus employed different model types.
We observed two broad categories of model types based on how they are obtained:

\begin{enumerate}[left=0.05em]
    \item Established Models (\mSC{establishedModel}) (75.94\%): Most workflows used well-established models like machine learning models (e.g., regression~\cite{koyama2014crowd,dingen2018regressionexplorer,gathani2021augmenting}, GAMs~\cite{hohman2019gamut,wang2023gam}, neural networks~\cite{park2021vatun,wang2021visual}, SVMs~\cite{gomez2021advice,xu2018ensemblelens,espadoto2021unprojection}), deep learning models (e.g., CNNs~\cite{li2020cnnpruner,park2021vatun,hamid2019visual}, BERT~\cite{bian2021semantic,liu2024visualizing}, sequence-to-sequence~\cite{strobelt2018s}), optimization models (e.g., SEIR~\cite{rydow2022development, tariq2021planning}, Monte Carlo Simulations~\cite{booshehrian2012vismon,tariq2021planning}), dimensionality reduction models (e.g., PCA~\cite{cavallo2018visual,espadoto2021unprojection,orban2018drag}, MDA~\cite{orban2018drag}), and domain-specific models (e.g., WRF climate model~\cite{biswas2016visualization}).
    
    \item User-Defined Models (\mSC{userDefinedModel}) (15.04\% specified by authors, 0.75\% defined by end-users at runtime): A few works in the corpus used specially defined models to allow tailoring to prior knowledge about the relationships between variables.
    For example, in the marketing domain, Gathani et al.~\cite{gathani2022if} model the \textit{Impression} output variable as a linear combination of various input features; 2$\times$\textit{Paid Media Ads} + 1$\times$\textit{Website Visit} + 1$\times$\textit{Video Completion Rate}, while Gortler et al.~\cite{gortler2019uncertainty} demonstrate a non-linear user-defined model by applying a scaling factor to the input covariance matrix, resulting in complex transformations of the projection directions.
\end{enumerate}

Across the corpus, models were obtained through different sources: 35.34\% were trained entirely by the authors~\cite{hamid2019visual,wang2019designing}, 6.77\% combined author-trained and user-defined models~\cite{wu2014opinionflow,wang2021visual}, 24.81\% used fine-tuned pre-trained models~\cite{browne2020camera,shaikh2024rehearsal}, and 9.77\% employed algorithms newly developed within the systems~\cite{wang2024empirical,crisan2024exploring}.

\vspace{1.5mm}
\noindent \textbf{3.2.4. Model Parameters}

\mSC{modelParameters} are internal variables governing how the \model maps input variables to the output variables (e.g., \mSC{featureWeights} and \mSC{biases} in machine learning models, and \mSC{kernels}, \mSC{attentionWeights}, and \mSC{positionalEncoding} in deep learning models).
In several workflows, these parameters are explicitly exposed as targets of interaction operations, enabling analysts to reason about hypothetical changes within the model itself in addition to variations in the input data.
For instance, Görtler et al.~\cite{gortler2019uncertainty} vary a \mSC{scalingFactor} that affects covariance in PCA, while Zhang et al.~\cite{zhang2022promotionlens} expose learned \mSC{featureWeights} to help analysts rank influential promotions.
\subsection{Interaction Operations}      
\label{subsec:operations}
Interaction operations are the third primitive class of \praxa.
Interaction operations describe \textit{how} the workflow is carried out.
They represent the sequence of user actions and corresponding system responses through which what-if workflows unfold over the \dataset and \model primitives.
Each interaction operation therefore consists of a pair: a \userOperations, initiated by the user and a corresponding \systemOperations executed by the system in response.
From our review of workflows across the corpus, we identified six such user–system pairings.
The first two--\uOpSC{scope} $\to$ \sOpSC{transform} and \uOpSC{derive} $\to$ \sOpSC{expand} (marked with \textbf{*})--are general data analysis operations (observed in SQL querying and exploratory data analysis) that, while not exclusive to what-if analysis, often appear as preparatory steps that establish the analytical context for subsequent what-if operations.

\vspace{1.5mm}
\noindent \textbf{3.3.1. \uOpSC{Scope} $\to$ \sOpSC{Transform}}

The \uOpSC{scope} user action restricts the \dataset to a targeted subset—by \dSC{instances}, \dSC{features}, or both—that becomes the focus of the workflow.
In response, the system executes \sOpSC{transform}, slicing and dicing the dataset to contain only the scoped subset so downstream operations act only on relevant data.
Instance-level scoping is illustrated by Zhang et al.~\cite{zhang2022promotionlens}, who allow marketing analysts to select specific data points, and by Wu et al.~\cite{wu2023scattershot}, who cluster textual documents into semantic groups to examine model behavior within subsets.
Feature-level scoping is illustrated by Gathani et al.~\cite{gathani2022if}, who exclude economic indicators such as the \textit{Consumer Confidence Index} and \textit{Producer Price Index} because business users cannot influence them--thereby restricting the workflow to actionable variables.

\vspace{1.5mm}
\noindent \textbf{3.3.2. \uOpSC{Derive} $\to$ \sOpSC{Expand}}

The \uOpSC{derive} user action creates a new \mSC{inputVariable} or \mSC{outputVariable} from existing \dataset features through arithmetic or statistical transformations.
In response, the system executes \sOpSC{expand}, extending the dataset with the newly created feature and making it available for subsequent operations in the workflow.
For example, Zgraggen et al.~\cite{zgraggen2016tableur} derive \textit{Total} by subtracting \textit{Expenses} from \textit{Income}, while Ahn et al.~\cite{ahn2024interactive} introduce a \textit{Miscalibration} output variable via KL divergence to quantify algorithmic harm in a movie recommendation dataset.

\vspace{1.5mm}
\noindent \textbf{3.3.3. \uOpSC{Perturb} $\to$ \sOpSC{Rerun}}

The \uOpSC{perturb} user action modifies the values of \mSC{inputVariables} or \mSC{modelParameters} to examine their effect on the \mSC{outputVariable}.
In response, the system executes \sOpSC{rerun}, re-executing the model's inference pipeline with the updated inputs to produce revised predictions.
Perturbations may target input variables--for instance, Wang et al.~\cite{wang2024empirical} examine how reducing \textit{Price} by 10\% affects \textit{Sales}--or model parameters, like Liu et al.~\cite{liu2018nlize} modify the \textit{attention} parameter in a CNN to evaluate its impact on sentence-level inference.
The \sOpSC{rerun} response may also be triggered by other operations such as \uOpSC{scope} and \uOpSC{derive}, when those actions alter the data or feature space on which the model operates, ensuring predictions remain consistent with the current analytical context.

\vspace{1.5mm}
\noindent \textbf{3.3.4. \uOpSC{Constrain} $\to$ \sOpSC{Optimize}}

The \uOpSC{constrain} user action places restrictions on various elements of the workflow, transforming the analysis into a directed search problem.
Constraints can be applied to the \model itself--for example, enforcing monotonic relationships (e.g., \textit{Sales} should not increase beyond a price threshold~\cite{gathani2025if}) or fairness requirements (e.g., excluding \textit{Gender} from model mappings~\cite{bhattacharya2023directive})--or directly to variable values, restricting \mSC{inputVariables} or \mSC{outputVariable} to specified bounds or categories (\sOpSC{closeTo(referenceDataPoint)}), or objectives (\sOpSC{minimize}, \sOpSC{maximize}).

In response, the system executes \sOpSC{optimize}, searching the feasible input space to identify \mSC{inputVariable} configurations that satisfy the specified constraints.
Across the corpus, three optimization objectives were commonly observed: \sOpSC{maximize} the \mSC{outputVariable} (e.g., \textit{Sales}~\cite{gathani2022if,gathani2021augmenting}, \textit{Income}~\cite{wexler2019if}), \sOpSC{minimize} it (e.g., \textit{Risk of Diabetes}~\cite{bhattacharya2023directive}, \textit{Number of Infections}~\cite{tariq2021planning}), or reach a \sOpSC{target} value or approximate a reference point (e.g., keeping \textit{Precipitation} close to baseline~\cite{hazarika2023haiva}).

A special case of this operation pair observed was counterfactual reasoning, where the system identifies the minimal changes to \mSC{inputVariables} required to produce a different output~\cite{bhattacharya2023directive,hao2023timetuner,shaikh2024rehearsal}.

\vspace{1.5mm}
\noindent \textbf{3.3.5. \uOpSC{Reweight} $\to$ \sOpSC{Retrain}}

The \uOpSC{reweight} user action adjusts the relative importance of \mSC{inputVariables} within the \model by modifying their \mSC{featureWeights}, directing model predictions toward a desired outcome based on domain knowledge or user priorities.

In response, the system executes \sOpSC{retrain}, updating the model so that predictions remain consistent with the revised variable importance.
For example, Pajer et al.~\cite{pajer2016weightlifter} reweight \textit{Price}, \textit{Mileage}, and \textit{Guarantee Period} in a car ranking model, prompting retraining to reflect the updated ranking preferences.
Retraining may also be triggered when the underlying model changes (\mSC{modelType}).
For instance, Wexler et al.~\cite{wexler2019if} allowed analysts to support comparison across model types such as linear classifiers and neural networks.

Alternatively, when feature importance is inferred from an existing model rather than explicitly modified by the user, the system executes \sOpSC{computeFeatureWeights}, evaluating the model to derive feature contributions without retraining.

\vspace{1.5mm}
\noindent \textbf{3.3.6. \uOpSC{Compare} $\to$ \sOpSC{Generate}}

The \uOpSC{compare} user action selects multiple outcomes from previously executed what-if workflows for juxtaposition, evaluating trade-offs and relative impacts across configurations.
Unlike earlier user actions that operate directly on data or model primitives, \uOpSC{compare} operates on results produced by prior workflow steps, making it a higher-level operation.
In response, the system executes \sOpSC{generate}, producing a structured set of scenarios--each consisting of an \mSC{inputVariable} configuration and its corresponding \mSC{outputVariable} prediction--and aligning them for comparison.
For example, Bhattacharya et al.~\cite{bhattacharya2023directive} compare treatment recommendations across patients, while Evirgen et al.~\cite{evirgen2024text} contrast alternative prompts in text-to-image generation.
More broadly, scenarios may be generated by applying the same perturbation across different data segments (e.g., geographic regions~\cite{gathani2021augmenting}) or under varying constraints~\cite{wang2024empirical}.

\section{Operationalizing \praxa: The Specification Language}
\label{sec:psl}
\praxa formalizes what-if analysis as a compositional grammar, but does not by itself enable execution.
To operationalize \praxa in systems, we encode its primitives into \pslfullname (\psl), a JSON-based declarative specification language in which each property maps directly to a \praxa primitive.
\psl makes what-if workflows explicit, portable, and system-independent.
By encoding \praxa's primitives, it supports execution and translation across systems, independent of domain or implementation.
This provides a unified, grammar-based representation of what-if analysis not offered by existing declarative approaches.

\subsection{Language Properties}
\label{subsec:properties}

\psl encodes \praxa's three primitive classes into six properties, each mapping onto a specific primitive or primitive subcomponent.
We explain these via a \psl specification for the workflow drawn from Wexler et al.'s What-If Tool~\cite{wexler2019if}: ``Would the person (data point 446) get
the loan if \textit{Capital Gain} is changed from 3411 to 20000?''
It specifies 
a point-value \uOpSC{perturb} on a single \mSC{inputVariable} and a binary \mSC{outputVariable} scoped to a single instance.
We use this as a running example to explain the properties underneath:

\vspace{1.5mm}
\noindent \textbf{\dataset}: \textit{\praxa's \dataset primitive}

\noindent The \dataset property encodes the \dataset primitive, specifying the data over which the workflow operates.
It is retrievable as a CSV file or PostgreSQL database. 
Additional \mSC{derivedFeatures} may be specified here when the workflow requires features created via the \uOpSC{derive} interaction operation.
In the running example, the UCI Census dataset is specified, with features including \textit{Capital Gain}, \textit{Age}, \textit{Hours-Per-Week}, and \textit{Gender}, among others.

\vspace{1.5mm}
\noindent \textbf{\mSC{outputVariable}}: \textit{\praxa's \mSC{outputVariable} element of the \model primitive}

\noindent The \mSC{outputVariable} property encodes the \mSC{outputVariable} element of the \model primitive--the dependent variable the workflow reasons about.
In the running example, \textit{Income} (binary: $>$50K or $\leq$50K) is the output variable, operationalizing the loan approval question.

\vspace{1.5mm}
\noindent \textbf{\sOpSC{objective}}: \textit{\praxa's analytical intent + goal for the \mSC{outputVariable}}

\noindent The \sOpSC{objective} property jointly encodes the analytical intent motivating the workflow and the desired behavior for the \mSC{outputVariable}.

In the running example, the analytical intent is to interpret and debug the model, achieved by understanding how perturbations to \textit{Capital Gain} affect \textit{Income} predictions and reveal when and why the model changes its output.
The goal for the \mSC{outputVariable} is to \sOpSC{set} it to a specific target value (e.g., $\geq$ \$50K).
Other \mSC{outputVariable} goals supported by \psl include \sOpSC{maximize} (e.g., \textit{Sales}) and \sOpSC{minimize} (e.g., \textit{Churn}).

\vspace{1.5mm}
\noindent \textbf{\model}: \textit{\praxa's \model primitive}

\noindent The \model property encodes the \model primitive, defining the predictive mechanism that connects \mSC{inputVariables} to the \mSC{outputVariable}.
Each entry specifies a unique \code{id}, \mSC{modelType}, and optional configuration parameters.
\psl may support many \mSC{modelTypes} like linear regressor, random forest regressor, and multi-layer perceptron regressor for quantitative output variables, and their classifier counterparts for nominal or ordinal output variables, as well as \mSC{userDefinedModel} entries for workflows requiring custom predictive mechanisms.
Multiple models can be specified to support workflows that compare predictions across \mSC{modelType} values, as in Wexler et al.~\cite{wexler2019if}.
In the running example, a \code{randomForestClassifier} is used since \textit{Income} is a binary variable.

\vspace{1.5mm}
\noindent \textbf{\uOpSC{scope}} \textit{(optional)}: \textit{\praxa's \uOpSC{scope} $\to$ \sOpSC{transform} interaction operation}

\noindent The \uOpSC{scope} property encodes the \uOpSC{scope} user action and its triggered \sOpSC{transform} system response, restricting the \dataset to a targeted subset before the other interaction operations are executed.
Each scope object carries a unique \code{id} for referencing different scopes to execute on the experiments.
Scope can be defined at the \code{instance} level--as in the running example, where the analysis is restricted to instance 446--or at the \code{feature} level, such as excluding \textit{Gender} from the \mSC{inputVariables} considered.
Specifying a scope is optional; its absence implies that the what-if configuration is applied on the full \dataset.

\vspace{1.5mm}
\noindent \textbf{\code{experiments}}: \textit{\praxa's interaction operations}

\noindent The \code{experiments} property encodes \praxa's third primitive class of the interaction operations via which a what-if workflow is carried out.
Each experiment object represents a single interaction step, defined as a paired user action and system response operating over the \dataset and \model primitives.
It references a \model and optionally a \uOpSC{scope} by their \code{id}, and specifies the user action to be performed (e.g., \uOpSC{perturb}, \uOpSC{constrain}, \uOpSC{reweight}, or \uOpSC{compare}).
The corresponding system response (e.g., \sOpSC{rerun}, \sOpSC{optimize}, \sOpSC{retrain} or \sOpSC{computeFeatureWeights}, \sOpSC{generate}) is implied by the pairing defined in \praxa.

Each user action is parameterized through operation-specific properties. For example, \code{perturb} specifies a target \mSC{inputVariable} or \mSC{modelParameter} along with a value or range, while \code{constrain} defines bounds and an optimization objective.
When operations act on prior results rather than directly on data or model primitives, \code{useOutputFrom} enables referencing outputs from earlier experiments.

Multiple experiment objects may be defined within a specification, each encoding a single interaction step, and together, they form a sequence constituting the overall workflow.

In the running example, a single experiment encodes a \uOpSC{perturb} action on \textit{Capital Gain}, changing its value from 3411' to 20000' for the scoped instance (data point 446). 
This triggers \sOpSC{rerun}, producing an updated prediction of \textit{Income}.

\subsection{Compilation and Execution}

A \psl specification is compiled into an executable workflow by resolving dependencies between primitives (e.g., \dataset, \model, and interaction operations) and constructing an execution plan over the specified steps. 
Each experiment encodes a user action, with the corresponding system response determined by the interaction operation pairing defined in \praxa.

During execution, these steps are applied sequentially over the \dataset and \model, producing outputs such as updated predictions, optimized inputs, or generated scenarios.
Outputs can be consumed directly or referenced in subsequent experiments.

This execution model follows \praxa's compositional structure, where workflows are not predefined but emerge from the combination of interaction operations. 
In the following section, we leverage this property to demonstrate the expressiveness and composability of \praxa through concrete workflow instantiations, and its role as an executable foundation for next-generation what-if systems.
\section{Evaluating the \praxa Grammar}
\label{sec:evaluation}

Having defined \praxa and its corresponding \psl representation, we evaluate what the grammar enables along three dimensions. 
First, we demonstrate its \textit{expressiveness} by showing that a small set of canonical configurations captures the range of what-if workflows observed across prior systems (Section~\ref{subsec:expressiveness}). 
Second, we examine its \textit{composability}, showing how these configurations can be systematically combined to form more complex, multi-step workflows that are largely unsupported in existing systems (Section~\ref{subsec:composability}).
Finally, we illustrate \psl as an intermediate representation for natural-language what-if interfaces, showing how structured specifications enable reliable translation from NL queries to executable visual workflows (Section~\ref{subsec:nl}).

\begin{figure}[!b]
    \centering
    \vspace{-6mm}
    \includegraphics[width=\columnwidth]{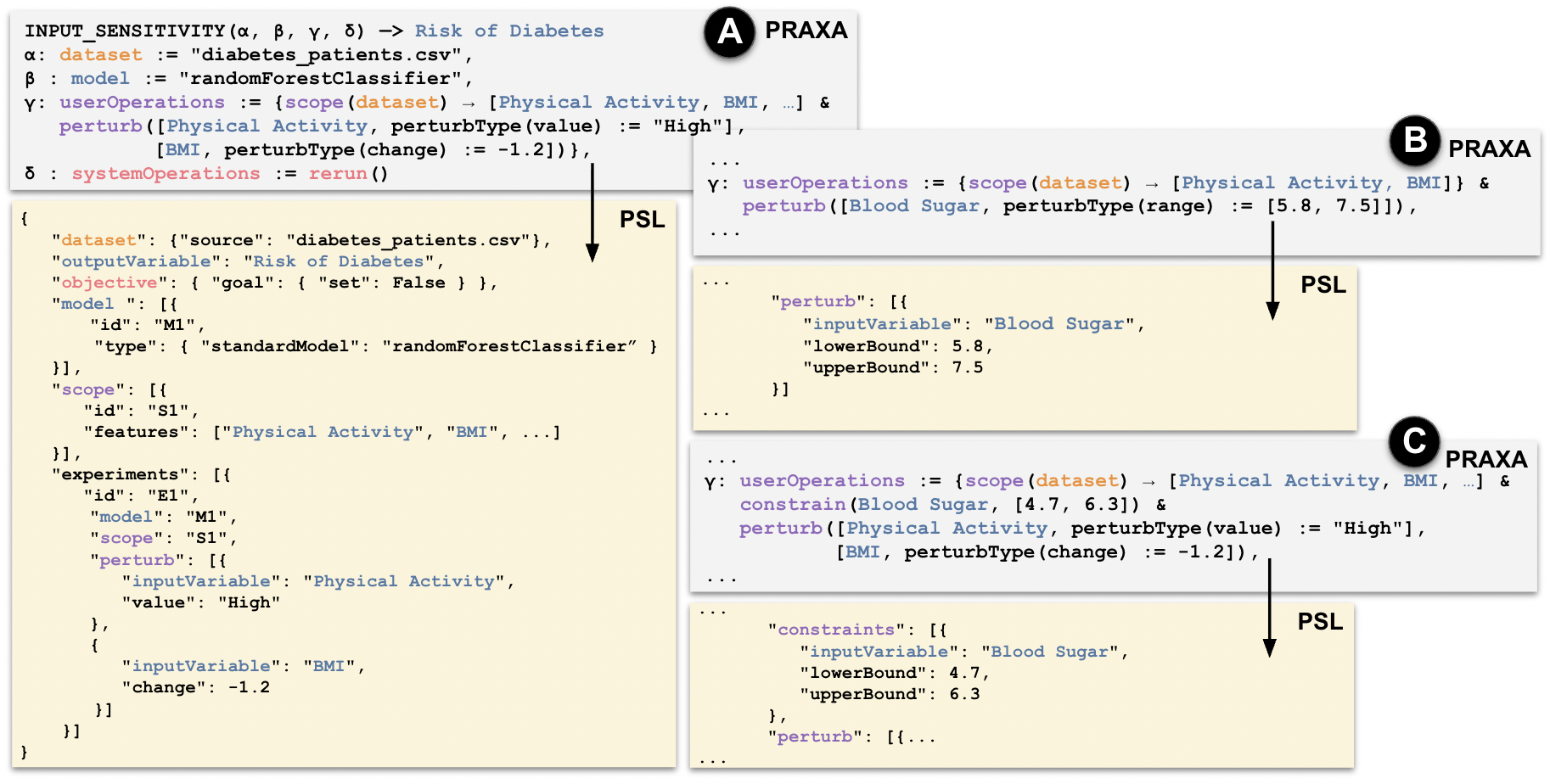}
    \vspace{-5mm}
    \caption{\praxa grammar and corresponding \psl encodings for \type{INPUT SENSITIVITY}, illustrating value-based (A), range-based (B), and constrained (C) perturbation configurations.}
    \label{fig:d1}
\end{figure}

\subsection{Expressiveness of Grammar}
\label{subsec:expressiveness}
\subsubsection{Canonical What-If Workflow Configurations}
\begin{figure*}[t]
    \centering
    \includegraphics[width=\textwidth]{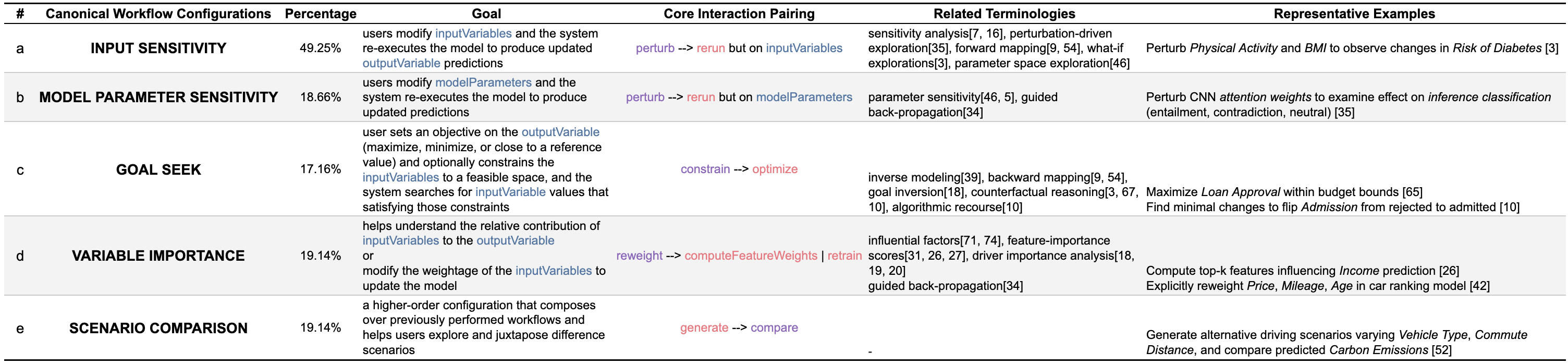}
    \vspace{-6mm}
    \caption{Five recurring workflow configurations expressed in \praxa, with their core interaction pairings, goals, related terminologies, and representative examples from the corpus.}
    \vspace{-5mm}
    \label{fig:expressiveness}
\end{figure*}

We demonstrate \praxa's expressiveness by showing that five canonical workflow configurations capture the range of what-if analyses observed across our corpus.
\autoref{fig:expressiveness} summarizes these configurations in terms of their interaction operation pairings, associated data and model primitives, related terminologies that instantiate the same structure, and representative systems from prior work.
For each configuration, we reconstruct the example systems as a composition of \praxa primitives and encode it as a \psl specification.
Each configuration is formalized as \type{CONFIGURATION($\alpha$, $\beta$, $\gamma$, $\delta$) $\rightarrow$ output}, where \codee{:=} denotes primitive instantiation.
We present both the grammar and corresponding \psl for the first configuration; for subsequent configurations, we emphasize key \psl components and defer full grammar specifications to the supplementary material~\cite{supplementary}.

\vspace{1.5mm}
\noindent \type{\textbf{a) INPUT SENSITIVITY.}}
This configuration captures workflows where users perturb \mSC{inputVariables} and the system recomputes the resulting \mSC{outputVariable} via the \uOpSC{perturb}(\mSC{inputVariables}) $\rightarrow$ \sOpSC{rerun} pairing. 
It is the most prevalent workflow in our corpus (49.25\%).

Despite being described under diverse terminologies (e.g., sensitivity analysis''~\cite{borland2024using,evirgen2024text}, perturbation-driven exploration''~\cite{liu2018nlize}, ``forward mapping''~\cite{splechtna2015interactive,cavallo2018visual}, etc.), they share the same underlying structure, differing in which variables are modified and how perturbations are specified.

For example, Bhattacharya et al.~\cite{bhattacharya2023directive} instantiate this workflow by allowing users to perturb actionable \mSC{inputVariables}--such as \textit{Physical Activity}, \textit{BMI}, and \textit{Blood Sugar}--and observe corresponding changes in predicted \textit{Risk of Diabetes} (\autoref{fig:d1}A).
Interactions are restricted to actionable features via a feature-level \uOpSC{scope}, excluding variables such as \textit{Age} and \textit{Gender}, and each perturbation triggers \sOpSC{rerun}, recomputing predictions under modified inputs.

Variations of this configuration include perturbing variables over continuous ranges (e.g., \textit{Blood Sugar} from `5.8–7.5'; \autoref{fig:d1}B) or composing perturbation with \uOpSC{constrain} (e.g., restricting \textit{Blood Sugar} to `4.7–6.3'; \autoref{fig:d1}C) to enforce domain-specific bounds, while preserving the same underlying structure.

\vspace{1.5mm}
\noindent \type{\textbf{b) MODEL PARAMETER SENSITIVITY.}}
This configuration captures workflows where users perturb \mSC{modelParameters} and the system recomputes outputs via the same \uOpSC{perturb} $\rightarrow$ \sOpSC{rerun} pairing as \type{INPUT SENSITIVITY}.
It is observed in 18.66\% of workflows and differs in its target: instead of modifying input data, it probes how changes to the model's internal configuration affect its behavior.

It appears under different terminologies like ``parameter sensitivity''~\cite{rydow2022development,biswas2016visualization} and ``guided back-propagation''~\cite{li2020cnnpruner}.

\begin{figure}[h]
    \centering
    \vspace{-3mm}
    \includegraphics[width=\columnwidth]{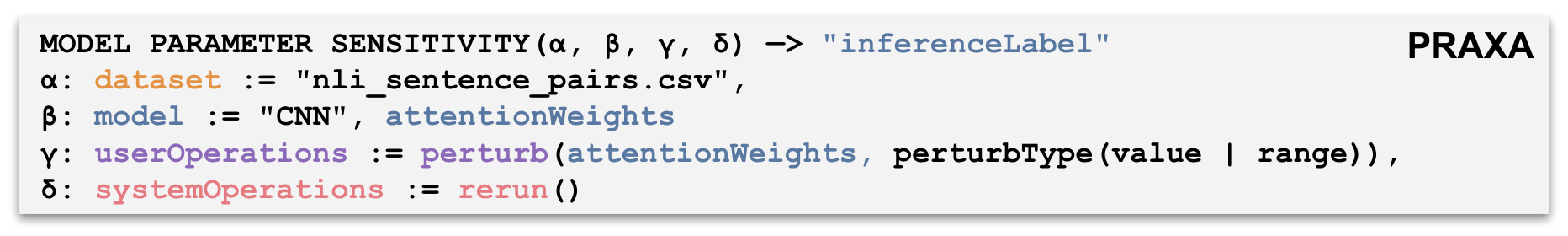}
    \vspace{-5mm}
    \caption{\praxa grammar for type{MODEL PARAMETER SENSITIVITY} workflow in Liu et al.'s NLIZE tool~\cite{liu2018nlize}.}
    \vspace{-3mm}
    \label{fig:d2}
\end{figure}

For example, Liu et al.~\cite{liu2018nlize} expose \textit{attention weights} in a CNN-based NL inference model as targets of perturbation (\autoref{fig:d2}). 
Users adjust which words or phrases the model attends to, triggering \sOpSC{rerun} to recompute classifications (e.g., \textit{entailment}, \textit{contradiction}, or \textit{neutral}).
This allows analysts to examine how internal weighting of linguistic features influences predictions, such as whether shifting attention to negation terms alters the inferred relationship.

\vspace{1.5mm}
\noindent \type{\textbf{c) GOAL SEEK.}}
Observed in 17.16\% of workflows, this configuration reverses the direction of sensitivity analysis: instead of observing how inputs affect outputs, users specify a desired outcome and the system determines input values that achieve it. 
Its core pairing is \uOpSC{constrain} $\rightarrow$ \sOpSC{optimize}, where users define objective on the \mSC{outputVariable} and optionally restricts \mSC{inputVariables} to a feasible space, and the system searches for input configurations satisfying those constraints.
The output includes both these \mSC{inputVariable} values and the resulting \mSC{outputVariable}.

This configuration appears under diverse terminologies (e.g., ``inverse modeling''~\cite{moon2023amortized}, ``backward mapping''~\cite{splechtna2015interactive,cavallo2018visual}, etc.), but shares the same underlying structure.

\begin{figure}[!b]
    \centering
    \vspace{-5mm}
    \includegraphics[width=\columnwidth]{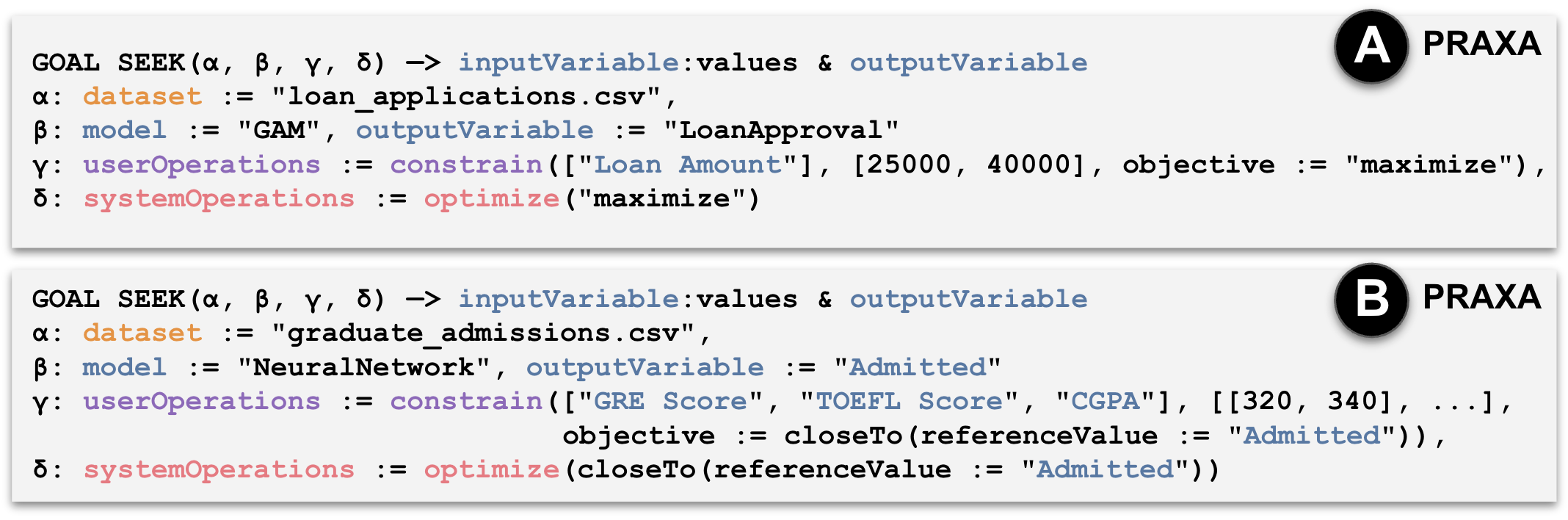}
    \vspace{-5mm}
    \caption{\praxa grammar for \type{GOAL SEEK}, illustrating maximize (A) and close to reference value (B) configurations.}
    \label{fig:d3}
\end{figure}

For example, Wang et al.~\cite{wang2023gam} instantiate this configuration in GAM Coach for algorithmic recourse (\autoref{fig:d3}A).
Given a model predicting \textit{Loan Approval}, users constrain actionable features such as \textit{Loan Amount} within feasible ranges (e.g., \$25000–\$40000) and set an objective to \codeee{maximize} approval probability. 
The system then optimizes over the constrained space to identify input configurations that achieve the highest likelihood of approval.

A closely related variation, often described as ``counterfactual reasoning'' across many works~\cite{bhattacharya2023directive,wexler2019if,cheng2020dece}, seeks minimal changes required to achieve a target outcome.
For instance, Cheng et al.~\cite{cheng2020dece} allow users to constrain features such as \textit{GRE Score}, \textit{TOEFL Score}, and \textit{CGPA} within realistic bounds and specify an objective of \codeee{closeTo("Admitted")} (\autoref{fig:d3}B).
The system searches for the nearest feasible input configuration that flips a prediction from `Rejected' to `Admitted'.

Across these cases, the underlying \uOpSC{constrain} $\rightarrow$ \sOpSC{optimize} structure remains the same, differing only in how the optimization objective is parameterized (e.g., \codeee{maximize} vs. proximity to a reference value), as shown in \autoref{fig:d3}.

As with other configurations, \uOpSC{scope} and \uOpSC{derive} may precede the core pairing.
The same structure also extends beyond tabular data--for example, Espadoto et al.~\cite{espadoto2021unprojection} implement inverse projection in dimensionality reduction by mapping 2D points back to high-dimensional representations via \sOpSC{optimize}, instantiating the same configuration at the level of data spaces rather than individual features.

\vspace{1.5mm}
\noindent \type{\textbf{d) VARIABLE IMPORTANCE.}} 
Observed in 19.40\% of workflows, this configuration focuses on understanding the contribution of \mSC{inputVariables} to the \mSC{outputVariable}. 
Its core interaction centers on \uOpSC{reweight}, but the corresponding system response depends on how importance is obtained. 
When importance is derived from an existing model, the system computes feature contributions (\sOpSC{computeFeatureWeights}) and when users explicitly modify feature weights, the system updates the model via \sOpSC{retrain}.

This configuration appears under diverse terminologies (e.g., ``influential factors''~\cite{xie2020visual,zhang2022promotionlens}, ``feature-importance scores''~\cite{laguna2023explimeable,hohman2019gamut,kaul2021improving}, and ``guided back-propagation''~\cite{li2020cnnpruner}), but shares the same underlying structure, differing only in whether \uOpSC{reweight} is implicit or explicit.

\begin{figure}[h]
    \centering
    \vspace{-3mm}
    \includegraphics[width=\columnwidth]{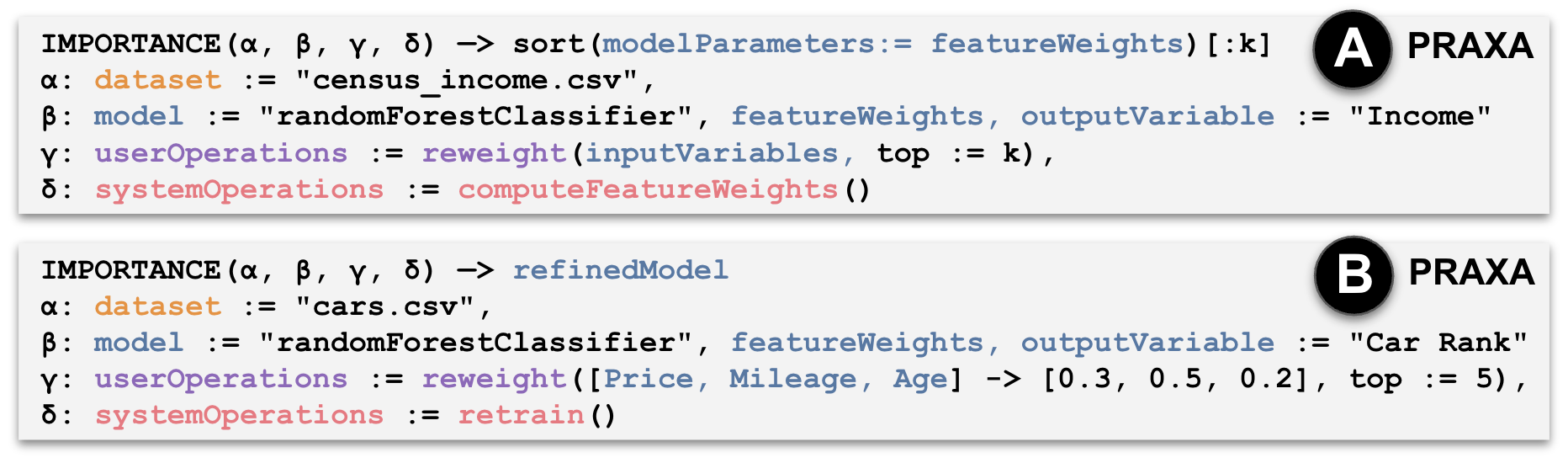}
    \vspace{-3mm}
    \caption{\praxa grammar for \type{IMPORTANCE}, illustrating getting the computed feature importances of \codeee{top-k} input variables (A) and \uOpSC{reweight} input variables to \sOpSC{retrain} the model (B) configurations.}
    \vspace{-3mm}
    \label{fig:d4}
\end{figure}

For example, Hohman et al.~\cite{hohman2019gamut} compute feature importance in a model predicting \textit{Income} from input variables such as \textit{Age}, \textit{Education}, and \textit{Hours-Per-Week}, producing ranked importance scores via \sOpSC{computeFeatureWeights} (\autoref{fig:d4}A).
In contrast, Pajer et al.~\cite{pajer2016weightlifter} allow users to explicitly assign weights to features such as \textit{Price}, \textit{Mileage}, and \textit{Age} in a car ranking model, triggering \sOpSC{retrain} to reflect updated preferences (\autoref{fig:d4}B).

As with other configurations, \uOpSC{scope} and \uOpSC{derive} may precede the core interaction.

\vspace{1.5mm}
\noindent \type{\textbf{e) SCENARIO COMPARISON.}}
Unlike the preceding configurations, this workflow operates over the outputs of prior analyses, enabling users to compare alternative scenarios and evaluate trade-offs.
It is observed in 19.40\% of workflows. 
Its core interaction is \sOpSC{generate} $\rightarrow$ \uOpSC{compare}, where the system produces a set of \mSC{inputVariable} configurations and corresponding \mSC{outputVariable} predictions, which users then juxtapose and interpret.

This configuration differs from others in two key ways.
First, the interaction operation's order is reversed: the system \sOpSC{generate}s results before user performs the \uOpSC{compare} operation. 
Second, it operates on previously computed outputs (e.g., from \type{GOAL SEEK}) rather than directly on data and model primitives, supporting open-ended exploration where the system generates a range of plausible scenarios, and the user compares and interprets alternatives.

\begin{figure}[t]
    \centering
    \vspace{-3mm}
    \includegraphics[width=\columnwidth]{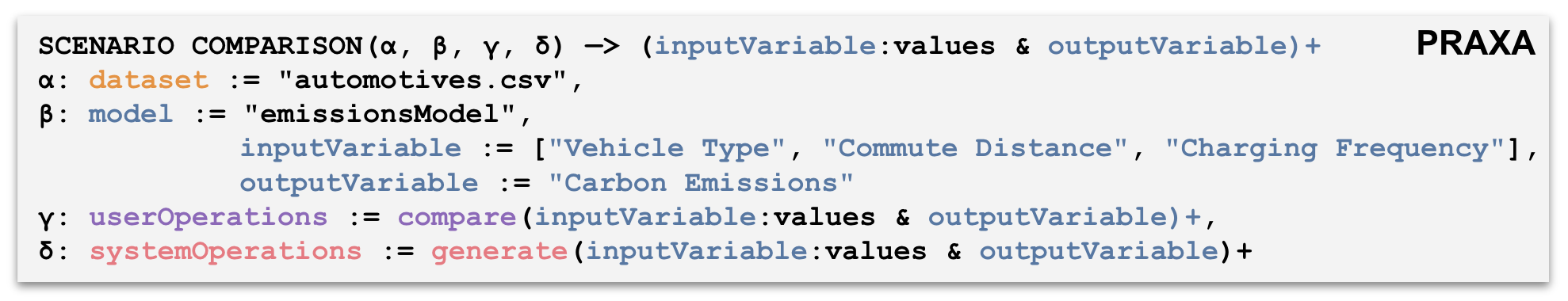}
    \caption{\praxa grammar for \type{SCENARIO COMPARISON} workflow for Shamma et al.'s EV Life dashboard~\cite{shamma2022ev}.}
    \vspace{-5mm}
    \label{fig:d5}
\end{figure}

For example, Shamma et al.~\cite{shamma2022ev} implement this configuration in EV Life, where the system generates alternative driving and ownership scenarios varying factors such as \textit{Vehicle Type} and \textit{Commute Distance}, along with predicted \textit{Carbon Emissions (\autoref{fig:d5})}. 
Users compare these scenarios side-by-side to assess which behavioral changes yield the greatest emission reductions.

\subsubsection{Findings from Existing Literature} 
\label{subhead:findings}
Mapping workflows in our corpus to \praxa's canonical configurations reveals two key insights about the current state of what-if analysis in visual analytics and HCI.

\noindent{\textbf{Terminological diversity and intent-based descriptions obscures shared workflow structure.}} 
Across the literature, what-if analysis is described through a wide range of domain- and task-specific terminologies, often presented as distinct analytical techniques. 
For example, approaches labeled as ``sensitivity analysis,'' ``perturbation-driven exploration,'' and ``forward mapping'' all reduce to the same \uOpSC{perturb} $\rightarrow$ \sOpSC{rerun} structure, differing only in whether perturbations target \mSC{inputVariables} or \mSC{modelParameters}. 
Similarly, ``inverse modeling,'' ``goal inversion,'' ``goal-seeking,'' and ``counterfactual reasoning'' all instantiate \uOpSC{constrain} $\rightarrow$ \sOpSC{optimize}, with variation arising primarily from how the optimization objective is parameterized.

Beyond terminology, analytical intent does not predict workflow structure. 
The same configuration can support multiple intents; for example, \type{INPUT SENSITIVITY} is used for both model debugging and evaluating intervention effects, while a single intent may be realized through different configurations (e.g., interpreting model behavior via \type{INPUT SENSITIVITY} or \type{IMPORTANCE}).
This lack of correspondence indicates that intent is orthogonal to workflow structure.

By organizing these approaches into canonical workflows, \praxa reveals these underlying equivalences and provides a consistent structural vocabulary independent of terminology or intent. 
More broadly, this suggests that many prior contributions introduce new parameterizations, instantiations, or domain interpretations of a smaller set of recurring structures rather than fundamentally new workflows.

\noindent{\textbf{Currently observed what-if workflows are predominantly single-step.}} 
Despite the diversity, most systems support only single-step workflows consisting of one user action followed by one corresponding system response (e.g., \uOpSC{perturb} $\rightarrow$ \sOpSC{rerun} or \uOpSC{constrain} $\rightarrow$ \sOpSC{optimize}).
Multi-step compositions--where outputs of one analysis are reused, refined, or compared in subsequent steps—are rarely supported. 
This reveals a gap between current system capabilities and the inherently compositional nature of what-if reasoning.
By making workflow structure explicit, \praxa shows that richer compositions are not exceptional cases but natural extensions of the same primitive building blocks.
\subsection{Composability and Reuse of Grammar}
\label{subsec:composability}

\subsubsection{Multi-Step Workflow Composition}
While existing systems predominantly support single-step workflows, what-if analysis in practice is inherently multi-step. 
Users rarely stop after a single analysis; instead, they compose multiple workflow configurations to achieve higher-level goals. 
For example, a user may identify an optimal solution via \type{GOAL SEEK}, evaluate its robustness via \type{INPUT SENSITIVITY}, and refine it further through repeated analysis.

\praxa enables such multi-step workflows through composition of interaction operations, where outputs from one configuration can be reused as inputs to another via \codeee{useOutputFrom}. 
This makes explicit the dataflow between workflows and exposes structural relationships that are otherwise hidden when workflows are treated as isolated techniques.

Using \praxa, we identify two recurring composition patterns that are analytically meaningful but largely unsupported in existing systems:

\vspace{1.5mm}
\subhead{a) Chained Workflows.} 
Chaining composes multiple canonical configurations sequentially, where the output of one workflow parameterizes the next.

\begin{figure}[h]
    \centering
    \vspace{-3mm}
    \includegraphics[width=\columnwidth]{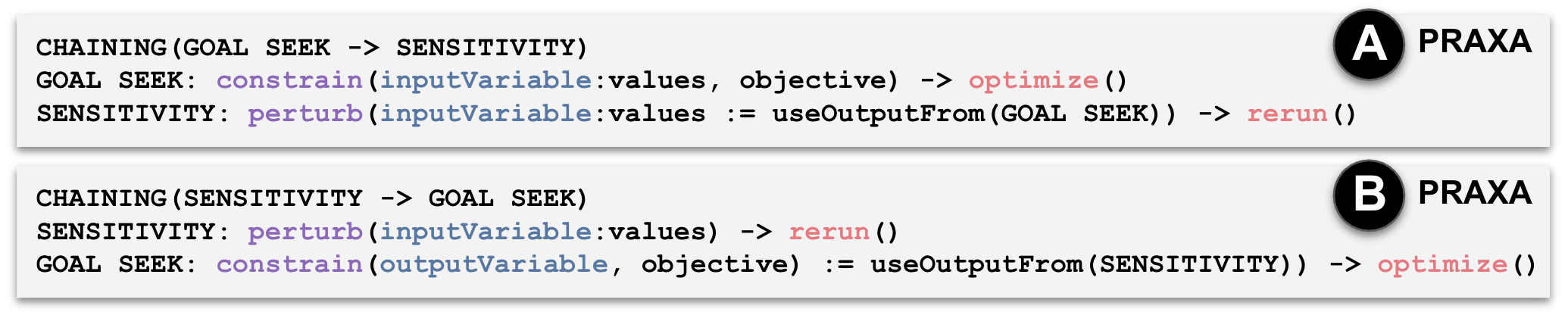}
    \vspace{-5mm}
    \caption{Chained what-if workflows expressed in \praxa. (A) \type{GOAL SEEK} $\rightarrow$ \type{INPUT SENSITIVITY}: optimized inputs are reused to evaluate robustness. (B) \type{INPUT SENSITIVITY} $\rightarrow$ \type{GOAL SEEK}: exploratory analysis informs subsequent optimization. Both compositions reuse intermediate outputs via \codeee{useOutputFrom}.}
    \vspace{-3mm}
    \label{fig:chain}
\end{figure}

\noindent \textit{a.1) \type{GOAL SEEK} $\rightarrow$ \type{INPUT SENSITIVITY} (optimize then validate).} 
A workflow first applies \uOpSC{constrain} $\rightarrow$ \sOpSC{optimize} to compute input configurations that achieve a desired outcome.
These computed values are then reused via \codeee{useOutputFrom} as the baseline for \uOpSC{perturb} $\rightarrow$ \sOpSC{rerun}, enabling robustness analysis under local variations (\autoref{fig:chain}A).

\noindent \textit{a.2) \type{INPUT SENSITIVITY} $\rightarrow$ \type{GOAL SEEK} (explore then target).} 
The reverse sequence begins with \uOpSC{perturb} $\rightarrow$ \sOpSC{rerun} to characterize how inputs influence outputs. 
These observed relationships inform a subsequent \uOpSC{constrain} $\rightarrow$ \sOpSC{optimize} step, where the optimization objective is parameterized using outputs from the prior analysis (\autoref{fig:chain}B).

In both cases, chaining requires no new primitives: each step is a canonical configuration, and composition is achieved solely through linking them via \codeee{useOutputFrom}.
Despite this simplicity, none of the systems in our corpus support such workflows as integrated processes, requiring users to manually transfer intermediate results.

\begin{figure}[h]
    \centering
    \vspace{-3mm}
    \includegraphics[width=\columnwidth]
    {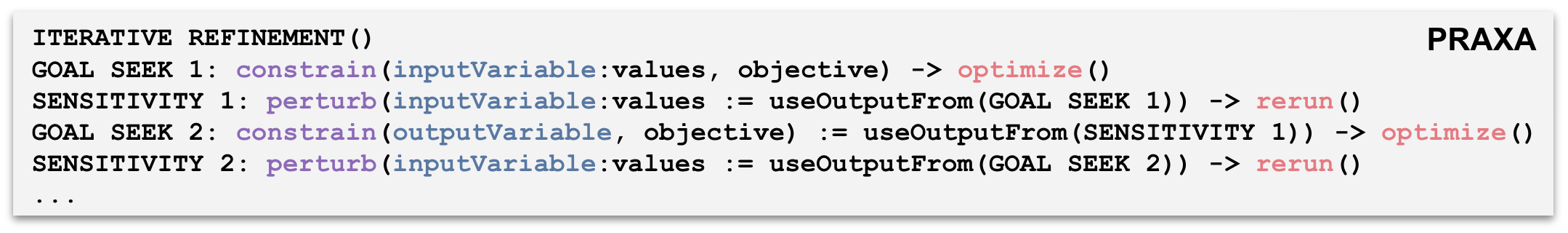}
    \vspace{-5mm}
    \caption{Iterative refinement as repeated composition of canonical workflows. Outputs from each step are reused to alternately optimize and validate solutions, enabling progressive refinement through \type{GOAL SEEK} and \type{INPUT SENSITIVITY}.}
    \vspace{-3mm}
    \label{fig:iterative_refinement}
\end{figure}

\subhead{b) Iterative Refinement.} 
Chained workflows naturally extend into iterative refinement, where users alternate between configurations to progressively improve both a solution and its robustness. 
As shown in \autoref{fig:iterative_refinement}, this pattern can be expressed as repeated compositions of \type{GOAL SEEK} and \type{INPUT SENSITIVITY}, where each step consumes outputs from the previous one.

This iterative structure reflects how analysts reason in practice: forming hypotheses, testing sensitivity, adjusting constraints, and repeating the process. 
However, current systems do not support maintaining or reusing intermediate states across iterations, forcing users to externalize this process.

These compositions do not introduce new analytical techniques; rather, they reflect workflows that users already perform implicitly by coordinating multiple tools and tracking intermediate results.
What is missing is system-level support for expressing and executing them as integrated processes. 
By formalizing multi-step workflows as compositions of primitives connected through \codeee{useOutputFrom}, \praxa provides the structure that existing systems lack and that next-generation what-if tools can build upon.

\subsubsection{Modification Cost Across Workflows} 
\label{subsec:modcost}
\praxa's compositional structure enables reasoning about structural relationships \textit{between} canonical workflows.
We define \textit{modification cost} as the number of primitive-level changes required to transform one configuration into another; how many primitives must be substituted, added, or removed while preserving the remaining structure.

Because all workflows share the same \dataset and \model primitives, structural distance is determined entirely by the interaction operations and their parameterization.

Several relationships are notable.
\type{INPUT SENSITIVITY} and \type{MODEL PARAMETER SENSITIVITY} are the closest pair (cost = 1), differing only in the target of perturbation—\mSC{inputVariables} versus \mSC{modelParameters}.
Transitioning from \type{INPUT SENSITIVITY} to \type{GOAL SEEK} requires changing both the user action (\uOpSC{perturb} $\rightarrow$ \uOpSC{constrain}) and system response (\sOpSC{rerun} $\rightarrow$ \sOpSC{optimize}), along with introducing an objective on the \mSC{outputVariable} (cost = 3).
\type{SCENARIO COMPARISON} is the most distant configuration (costs of 3--4), as it introduces a different interaction pairing (\uOpSC{compare} $\rightarrow$ \sOpSC{generate}) and operates over prior workflow outputs rather than directly on \dataset and \model primitives.
The exact cost depends on whether this shift is treated as a single structural change or decomposed into multiple primitive-level modifications.

These relationships indicate that canonical workflows occupy a shared structural space rather than representing isolated designs.
For system developers, this implies that supporting new workflows may require only localized changes to the interaction operation layer rather than redesigning data or model infrastructure.
For example, a system supporting \type{INPUT SENSITIVITY} (\uOpSC{perturb} $\rightarrow$ \sOpSC{rerun}) can support \type{GOAL SEEK} by introducing \uOpSC{constrain} $\rightarrow$ \sOpSC{optimize} over the same primitives.
\subsection{Illustrating \psl as an Intermediate Representation for NL What-if Queries}
\label{subsec:nl}
Beyond expressiveness and composability, \praxa's structure and explicit encoding to \psl creates a practical affordance: it makes what-if workflow specifications inspectable and correctable before execution.
This enables a potential application of \psl as an intermediate representation between NL what-if queries and executable interfaces.
Recent advances in Large Language Models (LLMs) have enabled NL interfaces for a wide range of data analysis tasks including SQL query generation~\cite{tian2024sqlucid,fu2023catsql,tian2025text}, chart generation~\cite{ko2024natural,tian2024chartgpt,luo2021natural}, and visual analytics~\cite{zhao2024leva,setlur2016eviza}.
Extending these capabilities to what-if analysis is challenging, as such workflows require structured reasoning over multiple primitives including data, models, and interaction operations. 
NL queries often under--specify this structure, making the intended analysis ambiguous.
\begin{figure}[!b]
    \centering
    \vspace{-3mm}
    \includegraphics[width=\columnwidth]{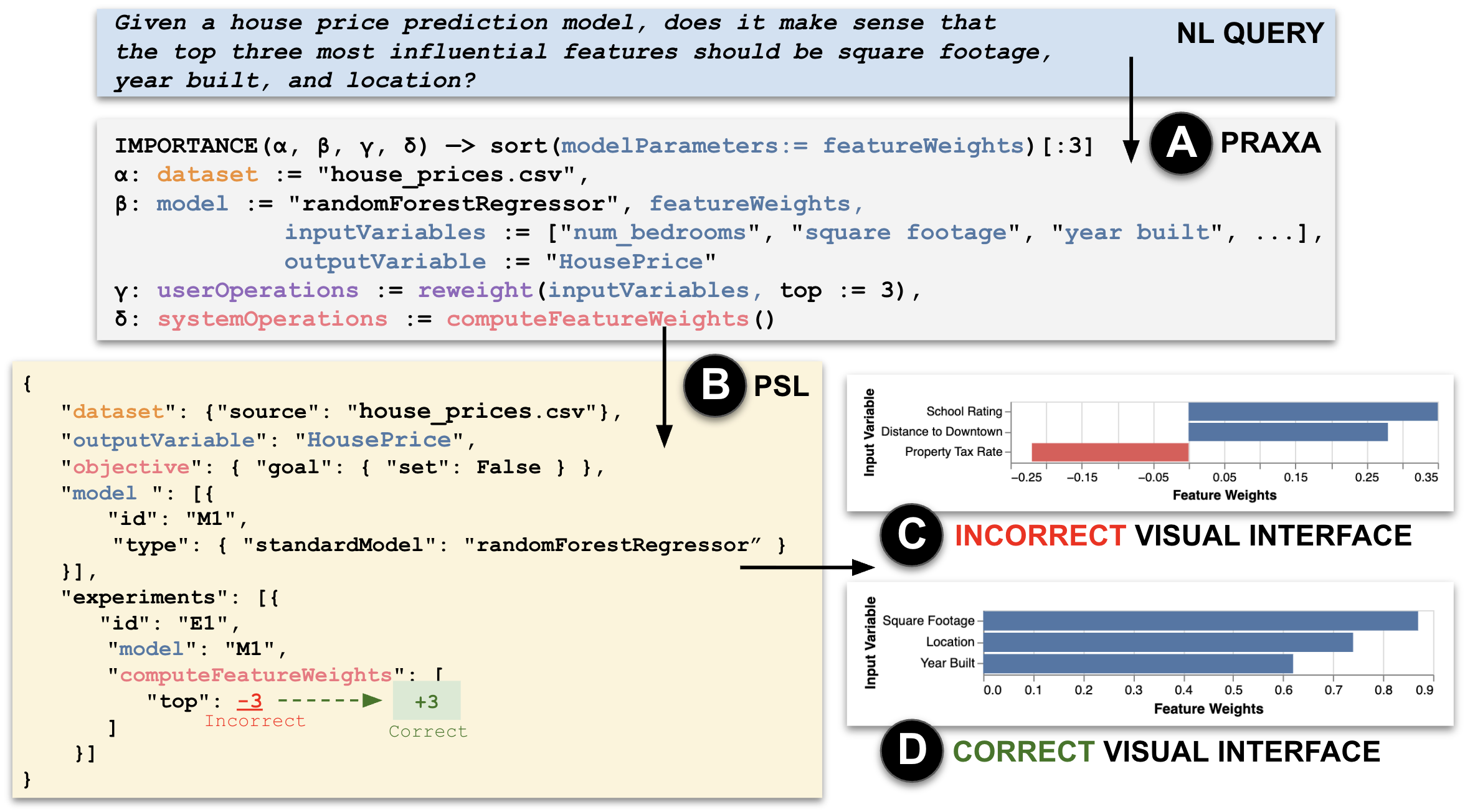}
    \vspace{-6mm}
    \caption{Illustrating \psl as an intermediate representation: a NL what-if query is composed into \praxa's grammar (A), encoded as a \psl specification (B), and compiled into an interactive visual interface. The explicit structure enables error detection--for instance, an incorrect ranking parameter (\codeee{"top": -3} vs. \codeee{"top": 3}) produces a plausible but incorrect interface (C) that can be identified and corrected in the specification before generating the intended interface (D).}
    \label{fig:nlwia}
\end{figure}
Systems that translate queries directly into code or interfaces may therefore produce outputs that appear plausible but misrepresent the user's analytical intent.

\psl addresses this by introducing a structured intermediate layer in which workflows are fully specified prior to execution. 
Queries are first translated into \psl specifications grounded in \praxa's primitives, making the analytical structure explicit and enabling validation before interface generation:

\subhead{NL Query to Interface.} 
To illustrate this property, consider the following NL query over a house price prediction model discussed by Hohman et al.~\cite{hohman2019gamut}: \textit{``Given a house price prediction model, does it make sense that the top three most influential features should be square footage, year built, and location?''} 
This corresponds to \type{IMPORTANCE} workflow, where the goal is to rank \mSC{inputVariables} by their influence on \mSC{outputVariable} (\textit{HousePrice}).
The query is first composed into \praxa's grammar (\autoref{fig:example}A), specifying the \dataset, \model, \mSC{outputVariable}, and an implicit \sOpSC{computeFeatureWeights} operation parameterized with \codeee{top := 3}.
This structure is then encoded as a \psl specification (\autoref{fig:example}B), where each property corresponds directly to a \praxa primitive. 
Compiling the correct specification produces an interface (\autoref{fig:example}D) that visualizes feature importance as a ranked bar chart.

\subhead{Error Detection through Structure.} 
The value of this intermediate layer becomes evident when translation errors occur. 
For instance, the phrase ``top three'' may be misinterpreted as the least influential features, appearing in \psl as \codeee{"top": -3} instead of \codeee{"top": 3}. 
Without an explicit specification, this error produces a plausible-looking visualization that silently displays the least influential features---answering the wrong question entirely (\autoref{fig:example}C vs.\ D).
With \psl, the error is localized to a single inspectable property, enabling straightforward correction prior to interface generation.

We emphasize that this example is illustrative rather than a fully realized system. 
We do not evaluate NL-to-\psl translation accuracy or robustness across queries.
Instead, the goal is to demonstrate that \praxa's compositional structure enables an explicit intermediate representation that supports inspection, validation, and repair--capabilities absent when NL queries are translated directly into interfaces. 
This highlights a promising direction for future NL-driven what-if systems.
\section{Discussion}
\label{sec:discussion}
Mapping the analyzed workflows form the corpus onto \praxa's canonical configurations reveals implications that extend beyond the grammar itself.
These findings point to gaps between the workflows supported by existing systems and the broader space of workflows that are structurally expressible.
We discuss four implications below, each highlighting opportunities for next-generation what-if analysis systems.

\subhead{The dominance of input sensitivity reflects evaluation practice.}
Nearly half of the workflows we analyzed in the corpus (49.25\%) implemented \type{INPUT SENSITIVITY}, the structurally simplest configuration.
In contrast, \type{GOAL SEEK}, which more directly supports decision-making and reasoning (i.e., identifying inputs that achieve a desired outcome), appeared in only 17.16\% of works.

This imbalance may be influenced by how what-if systems are currently evaluated and reported.
\type{INPUT SENSITIVITY} workflows are comparatively easier to implement, produce interpretable outputs, and lend themselves to controlled evaluation setups (e.g., varying a single input and observing changes in output).
In contrast, goal-seeking and multi-step workflows often involve search processes, multiple valid solutions, and more complex interpretation, making them harder to evaluate within standard experimental designs.

A similar pattern appears in the interaction and visualization designs of reviewed systems (described in more detail in the supplementary materials).
A substantial portion of works present predictive outputs without interactive manipulation~\cite{wang2019designing,evirgen2024text,hohman2019gamut}, while interactive systems frequently rely on simple controls such as sliders and table edits~\cite{bhattacharya2023directive,shen2024surroflow,wexler2019if}.
These design choices favor clarity and evaluability, but may limit the range of workflows that systems support.

\praxa makes this bias measurable by mapping systems onto canonical configurations.
This enables the community to track how supported workflows evolve over time and to better align system design and evaluation with the broader space of what-if analysis.

\subhead{Multi-step workflows are under-supported at the system level.}
The prevalence of single-step workflows observed in our corpus does not reflect a lack of need for multi-step what-if analysis.
In practice, users often combine multiple reasoning steps, for example, identifying an optimal strategy using \type{GOAL SEEK} and then assessing its robustness through \type{INPUT SENSITIVITY}, or alternating multiple single-step workflows continuously to come to a decision.

However, such multi-step workflows are rarely supported as integrated analytical processes in existing systems.
Our analysis suggests that this limitation stems from how current systems are structured.
Workflows are typically implemented as isolated operations, with limited support for storing, referencing, or composing intermediate results across steps.
As a result, users must manually coordinate multiple tools or repeat interactions to achieve what is conceptually a single reasoning process with multiple single-step configurations~\cite{tariq2021planning,zgraggen2016tableur}.

In contrast, \praxa makes multi-step workflows explicit as compositions of primitives, and \psl further operationalizes this by enabling workflows to reference prior results (e.g., via \code{useOutputFrom}), providing a concrete mechanism for chaining analyses.
This suggests a design direction in which what-if systems treat workflows as composable structures rather than isolated operations.

\subhead{Workflow structure remains implicit for users.}
A recurring challenge in prior work is that users struggle to decide what analyses to perform and how to configure them~\cite{pajer2016weightlifter,bhattacharya2023directive}.
Our findings suggest that this difficulty stems in part from the absence of explicit workflow structure.

Without a clear representation of analytical intents, interaction operations, and their compositions, users must rely on intuition to select variables, define constraints, and interpret results.
As workflows grow more complex--particularly in multi-step settings--this lack of structure increases cognitive load and makes it difficult to track how changes in inputs lead to outputs.

By formalizing what-if analysis as compositions of primitives, \praxa provides a foundation for making workflow structure explicit.
This opens opportunities for systems that guide users in selecting appropriate configurations, track analysis provenance, and support reproducible and iterative reasoning.

\subhead{Implications for NL what-if interfaces.}
Recent work has explored the use of NL interfaces for data analysis tasks such as visual analytics~\cite{zhao2024leva,setlur2016eviza}, chart generation~\cite{ko2024natural,tian2024chartgpt,luo2021natural}, and SQL querying~\cite{tian2024sqlucid,fu2023catsql,tian2025text}.
Extending such interfaces to what-if analysis introduces additional challenges due to the terminological fragmentation identified in this work and resonated by others~\cite{tariq2021planning,dhanoa2021fuzzy,zgraggen2016tableur}.
First, terminological fragmentation introduces structural ambiguity: the same terms (e.g., ``scenario analysis'') may map to different workflow configurations, while different terms (e.g., ``counterfactual reasoning'' and ``goal-seeking'') often share the same underlying structure.
Unlike SQL or visualization grammars, what-if queries lack a well-defined target representation, making correct translation inherently underspecified.
Second, what-if workflows are inherently multi-primitive, jointly involving data, models, and interaction operations.
Prior work shows that even expert users struggle to articulate this structure, often relying on ambiguous descriptions.
This is compounded by the predominance of single-step workflows in current systems, limiting both user expression and system inference of multi-step analyses.

These challenges highlight the need for structural grounding in NL what-if systems. As shown in \autoref{subsec:nl}, directly translating NL queries into interfaces can produce plausible but incorrect outputs when workflow structure is misinterpreted. Instead, systems can translate queries into compositional workflow representations like \praxa and \psl where primitives are explicit and inspectable.

\subsection{Limitations and Future Work}
\label{subsec:limitations}
First, \praxa was derived from a corpus of visual analytics and HCI publications and may not fully capture what-if workflows in adjacent fields such as operations research, decision science, or business intelligence, where different terminologies and structures may prevail. Extending and validating the grammar across these domains remains an important direction for future work.

Second, while \psl encodes \praxa's primitives into an executable specification language, the current implementation supports a limited set of model types and interaction patterns.
Expanding \psl's coverage--particularly to support more complex model architectures, uncertainty-aware predictions, and richer multi-step workflow compositions--remains open.

Third, the two applications demonstrated in this paper--dashboard compilation and NL translation, which demonstrate how \praxa can be operationalized, but do not constitute a formal evaluation of its usability and effectiveness in practice.
Future work should assess whether \praxa-grounded systems enable constructing and reasoning about more complex workflows, and whether the multi-step configurations identified in this work align with real analytical needs.

Finally, \praxa currently captures the structural composition of what-if workflows, but does not explicitly model the cognitive or collaborative needs of analysis, like how users decide which workflows to execute, how to iterate on them, collaborate around them, or communicate its results.
Extending \praxa to include an explanation layer that captures causal assumptions, uncertainty metadata, and provenance of workflow decisions would substantially broaden its utility as a foundation for user-adaptive what-if systems.
\vspace{-3mm}
\section{Related Work}
\label{sec:relatedWork}
Our work relates to prior research on what-if analysis support, visual parameter space analysis, and declarative specifications as intermediate representations.

\subhead{What-if Analysis Support.} What-if analysis has been studied as a critical capability for exploring hypothetical scenarios and supporting decision-making across domains.
Research systems have addressed WIA in specific contexts: Decision Studio~\cite{gathani2021augmenting} supports business professionals in configuring multiple what-if scenarios; Bhattacharya et al.~\cite{bhattacharya2023directive} prototype a patient-facing system for exploring diabetes risk through what-if explorations; and Tariq et al.~\cite{tariq2021planning} enable planning epidemic interventions through constrained optimization.
Gathani et al.~\cite{gathani2025if} provide a comprehensive study of how business professionals currently perform WIA in practice, identifying challenges and opportunities. However, these works each address WIA within specific domains and terminologies.
\praxa differs by deriving a domain-agnostic grammar from structural patterns across 141 publications, formalizing what these systems share rather than how they differ.

\subhead{Visual Parameter Space Analysis.} Several frameworks partially overlap with \praxa but differ in focus and scope.
Sedlmair et al.~\cite{sedlmair2014visual} outline high-level data flows, navigation strategies, and six tasks for parameter space analysis--\textit{optimization}, \textit{fitting}, \textit{sensitivity}, \textit{outliers}, \textit{uncertainty}, and \textit{partitioning}.
\praxa extends this work by providing a finer-grained compositional grammar that maps these abstractions to specific interaction operations and their compositions.
For instance, their \textit{optimization} and \textit{fitting} tasks correspond to our \type{GOAL SEEK} configuration (\uOpSC{constrain} $\rightarrow$ \sOpSC{optimize}), while \textit{sensitivity} aligns with our \type{INPUT SENSITIVITY} (\uOpSC{perturb} $\rightarrow$ \sOpSC{rerun}), and \textit{partitioning} relates to \type{SCENARIO COMPARISON}.
Where Sedlmair et al. emphasize task-level navigation, \praxa decomposes workflows into constituent primitives--data, model, and interaction operations--explaining how different components combine to produce structurally distinct analyses.

Piccolotto et al.~\cite{piccolotto2023visual} review visual parameter space exploration with a focus on how model parameters relate to outputs in spatial and temporal contexts.
While overlapping in sensitivity and optimization, their framework emphasizes high-level workflow actions (e.g., manual or automatic parameter setting).
\praxa is broader and domain-agnostic, positioning parameter space analysis as one instance of what-if analysis and decomposing it into composable primitives that generalize across analysis types.
Von et al.~\cite{von2016visualization} review visual analytics for optimizing structured algorithmic pipelines through application case studies, while La et al.~\cite{la2023state} survey visual analytics for explainable deep learning, recognizing perturbation and counterfactual techniques.
\praxa extends beyond both by systematically identifying the structural components shared across these techniques and formalizing them into a grammar derived from a larger corpus.
\vspace{-3mm}
\section{Conclusion}
We presented \praxa, a compositional grammar of what-if analysis derived from a structural analysis of 141 visual analytics and HCI publications. 
By formalizing workflows as compositions of three primitive classes--data, model, and interaction operations--\praxa reveals that what-if analyses in current works are described under diverse terminologies but share common structure and that existing systems are largely limited to single-step interactions despite users' multi-step reasoning.
We demonstrated the grammar's composability through chaining and iterative refinement, and illustrated it via \psl as an intermediate representation for translating NL queries into interactive interfaces, enabling structural validation and error detection.

\maketitle

\bibliographystyle{abbrv-doi-hyperref}

\bibliography{template}

@inproceedings{metz2024interactive,
  title={Interactive Public Transport Infrastructure Analysis through Mobility Profiles: Making the Mobility Transition Transparent},
  author={Metz, Yannick and Ackermann, Dennis and Keim, Daniel A and Fischer, Maximilian T},
  booktitle={2024 IEEE Visualization in Data Science (VDS)},
  pages={6--14},
  year={2024},
  organization={IEEE}
}

@inproceedings{levesque2024pathways,
  title={Pathways Explorer: Interactive Visualization of Climate Transition Scenarios},
  author={L{\'e}vesque, Fran{\c{c}}ois and Beaumier, Louis and Hurtut, Thomas},
  booktitle={2024 IEEE Workshop on Energy Data Visualization (EnergyVis)},
  pages={29--33},
  year={2024},
  organization={IEEE}
}

@article{de2023prowis,
  title={PROWIS: A Visual Approach for Building, Managing, and Analyzing Weather Simulation Ensembles at Runtime},
  author={de Souza, Carolina Veiga Ferreira and Bonnet, Suzanna Maria and de Oliveira, Daniel and Cataldi, Marcio and Miranda, Fabio and Lage, Marcos},
  journal={IEEE Transactions on Visualization and Computer Graphics},
  year={2023},
  publisher={IEEE}
}

@inproceedings{laguna2023explimeable,
  title={ExpLIMEable: A Visual Analytics Approach for Exploring LIME},
  author={Laguna, Sonia and Heidenreich, Julian N and Sun, Jiugeng and Cetin, Nil{\"u}fer and Al-Hazwani, Ibrahim and Schlegel, Udo and Cheng, Furui and El-Assady, Mennatallah},
  booktitle={2023 Workshop on Visual Analytics in Healthcare (VAHC)},
  pages={27--33},
  year={2023},
  organization={IEEE}
}

@article{li2020cnnpruner,
  title={Cnnpruner: Pruning convolutional neural networks with visual analytics},
  author={Li, Guan and Wang, Junpeng and Shen, Han-Wei and Chen, Kaixin and Shan, Guihua and Lu, Zhonghua},
  journal={IEEE Transactions on Visualization and Computer Graphics},
  volume={27},
  number={2},
  pages={1364--1373},
  year={2020},
  publisher={IEEE}
}

@ARTICLE{sarikaya2019what,
  author={Sarikaya, Alper and Correll, Michael and Bartram, Lyn and Tory, Melanie and Fisher, Danyel},
  journal={IEEE Transactions on Visualization and Computer Graphics}, 
  title={What Do We Talk About When We Talk About Dashboards?}, 
  year={2019},
  volume={25},
  number={1},
  pages={682-692},
  doi={10.1109/TVCG.2018.2864903}
}

@inproceedings{lee2022sleepguru,
  title={SleepGuru: Personalized Sleep Planning System for Real-life Actionability and Negotiability},
  author={Lee, Jungeun and Kim, Sungnam and Cheon, Minki and Ju, Hyojin and Lee, JaeEun and Hwang, Inseok},
  booktitle={Proceedings of the 35th Annual ACM Symposium on User Interface Software and Technology},
  pages={1--16},
  year={2022}
}

@article{seebacher2021investigating,
  title={Investigating the Sketchplan: A novel way of identifying tactical behavior in massive soccer datasets},
  author={Seebacher, Daniel and Polk, Tom and Janetzko, Halldor and Keim, Daniel A and Schreck, Tobias and Stein, Manuel},
  journal={IEEE Transactions on Visualization and Computer Graphics},
  volume={29},
  number={4},
  pages={1920--1936},
  year={2021},
  publisher={IEEE}
}

@inproceedings{wu2021principles,
  title={Principles and Interactive Tools for Evaluating and Improving the Behavior of Natural Language Processing models},
  author={Wu, Tongshuang},
  booktitle={Extended Abstracts of the 2021 CHI Conference on Human Factors in Computing Systems},
  pages={1--6},
  year={2021}
}

@article{kaul2021improving,
  title={Improving visualization interpretation using counterfactuals},
  author={Kaul, Smiti and Borland, David and Cao, Nan and Gotz, David},
  journal={IEEE Transactions on Visualization and Computer Graphics},
  volume={28},
  number={1},
  pages={998--1008},
  year={2021},
  publisher={IEEE}
}

@article{gathani2022if,
  title={What-if Analysis for Business Users: Current Practices and Future Opportunities},
  author={Gathani, Sneha and Liu, Zhicheng and Haas, Peter J and Demiralp, {\c{C}}a{\u{g}}atay},
  journal={2025 CHI Conference on Human Factors in Computing Systems},
  pages={arXiv--2212},
  year={2025}
}

@article{orban2018drag,
  title={Drag and track: A direct manipulation interface for contextualizing data instances within a continuous parameter space},
  author={Orban, Daniel and Keefe, Daniel F and Biswas, Ayan and Ahrens, James and Rogers, David},
  journal={IEEE transactions on visualization and computer graphics},
  volume={25},
  number={1},
  pages={256--266},
  year={2018},
  publisher={IEEE}
}

@inproceedings{moon2023amortized,
  title={Amortized inference with user simulations},
  author={Moon, Hee-Seung and Oulasvirta, Antti and Lee, Byungjoo},
  booktitle={Proceedings of the 2023 CHI Conference on Human Factors in Computing Systems},
  pages={1--20},
  year={2023}
}

@inproceedings{raidou2016visual,
  title={Visual analysis of tumor control models for prediction of radiotherapy response},
  author={Raidou, Renata Georgia and Casares-Magaz, Oscar and Muren, Ludvig P and Van der Heide, Uulke A and R{\o}rvik, Jarle and Breeuwer, Marcel and Vilanova, Anna},
  booktitle={Computer Graphics Forum},
  volume={35},
  number={3},
  pages={231--240},
  year={2016},
  organization={Wiley Online Library}
}

@article{gathani2021augmenting,
  title={Augmenting decision making via interactive what-if analysis},
  author={Gathani, Sneha and Hulsebos, Madelon and Gale, James and Haas, Peter J and Demiralp, {\c{C}}a{\u{g}}atay},
  journal={arXiv preprint arXiv:2109.06160},
  year={2021}
}

@inproceedings{bhattacharya2023directive,
  title={Directive explanations for monitoring the risk of diabetes onset: introducing directive data-centric explanations and combinations to support what-if explorations},
  author={Bhattacharya, Aditya and Ooge, Jeroen and Stiglic, Gregor and Verbert, Katrien},
  booktitle={Proceedings of the 28th international conference on intelligent user interfaces},
  pages={204--219},
  year={2023}
}

@inproceedings{waser2014many,
  title={Many plans: Multidimensional ensembles for visual decision support in flood management},
  author={Waser, J{\"u}rgen and Konev, Artem and Sadransky, Bernhard and Horv{\'a}th, Zsolt and Ribi{\v{c}}i{\'c}, Hrvoje and Carnecky, Robert and Kluding, Patrick and Schindler, Benjamin},
  booktitle={Computer Graphics Forum},
  volume={33},
  number={3},
  pages={281--290},
  year={2014},
  organization={Wiley Online Library}
}

@inproceedings{evirgen2024text,
  title={From text to pixels: Enhancing user understanding through text-to-image model explanations},
  author={Evirgen, Noyan and Wang, Ruolin and Chen, Xiang'Anthony},
  booktitle={Proceedings of the 29th International Conference on Intelligent User Interfaces},
  pages={74--87},
  year={2024}
}

@article{borland2024using,
  title={Using counterfactuals to improve causal inferences from visualizations},
  author={Borland, David and Wang, Arran Zeyu and Gotz, David},
  journal={IEEE Computer Graphics and Applications},
  volume={44},
  number={1},
  pages={95--104},
  year={2024},
  publisher={IEEE}
}

@article{liu2018nlize,
  title={Nlize: A perturbation-driven visual interrogation tool for analyzing and interpreting natural language inference models},
  author={Liu, Shusen and Li, Zhimin and Li, Tao and Srikumar, Vivek and Pascucci, Valerio and Bremer, Peer-Timo},
  journal={IEEE transactions on visualization and computer graphics},
  volume={25},
  number={1},
  pages={651--660},
  year={2018},
  publisher={IEEE}
}

@inproceedings{shamma2022ev,
  title={EV Life: A Counterfactual Dashboard Towards Reducing Carbon Emissions of Automotive Behaviors},
  author={Shamma, David A and Lee, Matthew L and Filipowicz, Alexandre LS and Denoue, Laurent and Glazko, Kate and Murakami, Kalani and Lyons, Kent},
  booktitle={Companion Proceedings of the 27th International Conference on Intelligent User Interfaces},
  pages={46--49},
  year={2022}
}

@article{sedlmair2014visual,
  title={Visual parameter space analysis: A conceptual framework},
  author={Sedlmair, Michael and Heinzl, Christoph and Bruckner, Stefan and Piringer, Harald and M{\"o}ller, Torsten},
  journal={IEEE Transactions on Visualization and Computer Graphics},
  volume={20},
  number={12},
  pages={2161--2170},
  year={2014},
  publisher={IEEE}
}

@article{wexler2019if,
  title={The what-if tool: Interactive probing of machine learning models},
  author={Wexler, James and Pushkarna, Mahima and Bolukbasi, Tolga and Wattenberg, Martin and Vi{\'e}gas, Fernanda and Wilson, Jimbo},
  journal={IEEE transactions on visualization and computer graphics},
  volume={26},
  number={1},
  pages={56--65},
  year={2019},
  publisher={IEEE}
}

@inproceedings{la2023state,
  title={State of the art of visual analytics for explainable deep learning},
  author={La Rosa, Biagio and Blasilli, Graziano and Bourqui, Romain and Auber, David and Santucci, Giuseppe and Capobianco, Roberto and Bertini, Enrico and Giot, Romain and Angelini, Marco},
  booktitle={Computer Graphics Forum},
  volume={42},
  number={1},
  pages={319--355},
  year={2023},
  organization={Wiley Online Library}
}

@inproceedings{hamid2019visual,
  title={Visual Ensemble Analysis to Study the Influence of Hyper-parameters on Training Deep Neural Networks.},
  author={Hamid, Sagad and Derstroff, Adrian and Klemm, S{\"o}ren and Ngo, Quynh Quang and Jiang, Xiaoyi and Linsen, Lars},
  booktitle={MLVis@ EuroVis},
  pages={19--23},
  year={2019}
}

@article{rydow2022development,
  title={Development and evaluation of two approaches of visual sensitivity analysis to support epidemiological modeling},
  author={Rydow, Erik and Borgo, Rita and Fang, Hui and Torsney-Weir, Thomas and Swallow, Ben and Porphyre, Thibaud and Turkay, Cagatay and Chen, Min},
  journal={IEEE Transactions on Visualization and Computer Graphics},
  volume={29},
  number={1},
  pages={1255--1265},
  year={2022},
  publisher={IEEE}
}

@inproceedings{park2021vatun,
  title={VATUN: Visual Analytics for Testing and Understanding Convolutional Neural Networks.},
  author={Park, Cheonbok and Yang, Soyoung and Na, Inyoup and Chung, Sunghyo and Shin, Sungbok and Kwon, Bum Chul and Park, Deokgun and Choo, Jaegul},
  booktitle={EuroVis (Short Papers)},
  pages={7--11},
  year={2021}
}

@article{gortler2019uncertainty,
  title={Uncertainty-aware principal component analysis},
  author={G{\"o}rtler, Jochen and Spinner, Thilo and Streeb, Dirk and Weiskopf, Daniel and Deussen, Oliver},
  journal={IEEE Transactions on Visualization and Computer Graphics},
  volume={26},
  number={1},
  pages={822--831},
  year={2019},
  publisher={IEEE}
}

@article{wang2021visual,
  title={Visual analytics for rnn-based deep reinforcement learning},
  author={Wang, Junpeng and Zhang, Wei and Yang, Hao and Yeh, Chin-Chia Michael and Wang, Liang},
  journal={IEEE Transactions on Visualization and Computer Graphics},
  volume={28},
  number={12},
  pages={4141--4155},
  year={2021},
  publisher={IEEE}
}

@article{wu2014opinionflow,
  title={Opinionflow: Visual analysis of opinion diffusion on social media},
  author={Wu, Yingcai and Liu, Shixia and Yan, Kai and Liu, Mengchen and Wu, Fangzhao},
  journal={IEEE transactions on visualization and computer graphics},
  volume={20},
  number={12},
  pages={1763--1772},
  year={2014},
  publisher={IEEE}
}

@inproceedings{koyama2014crowd,
  title={Crowd-powered parameter analysis for visual design exploration},
  author={Koyama, Yuki and Sakamoto, Daisuke and Igarashi, Takeo},
  booktitle={Proceedings of the 27th annual ACM symposium on User interface software and technology},
  pages={65--74},
  year={2014}
}

@inproceedings{shaikh2024rehearsal,
  title={Rehearsal: Simulating conflict to teach conflict resolution},
  author={Shaikh, Omar and Chai, Valentino Emil and Gelfand, Michele and Yang, Diyi and Bernstein, Michael S},
  booktitle={Proceedings of the 2024 CHI Conference on Human Factors in Computing Systems},
  pages={1--20},
  year={2024}
}

@inproceedings{wu2023scattershot,
  title={Scattershot: Interactive in-context example curation for text transformation},
  author={Wu, Sherry and Shen, Hua and Weld, Daniel S and Heer, Jeffrey and Ribeiro, Marco Tulio},
  booktitle={Proceedings of the 28th International Conference on Intelligent User Interfaces},
  pages={353--367},
  year={2023}
}

@article{biswas2016visualization,
  title={Visualization of time-varying weather ensembles across multiple resolutions},
  author={Biswas, Ayan and Lin, Guang and Liu, Xiaotong and Shen, Han-Wei},
  journal={IEEE transactions on visualization and computer graphics},
  volume={23},
  number={1},
  pages={841--850},
  year={2016},
  publisher={IEEE}
}

@inproceedings{wang2019designing,
  title={Designing theory-driven user-centric explainable AI},
  author={Wang, Danding and Yang, Qian and Abdul, Ashraf and Lim, Brian Y},
  booktitle={Proceedings of the 2019 CHI conference on human factors in computing systems},
  pages={1--15},
  year={2019}
}

@inproceedings{ahn2024interactive,
  title={Interactive Counterfactual Exploration of Algorithmic Harms in Recommender Systems},
  author={Ahn, Yongsu and Wolter, Quinn K and Dick, Jonilyn and Dick, Janet and Lin, Yu-Ru},
  booktitle={2024 IEEE Visualization in Data Science (VDS)},
  pages={35--39},
  year={2024},
  organization={IEEE}
}

@inproceedings{tariq2021planning,
  title={Planning epidemic interventions with EpiPolicy},
  author={Tariq, Zain and Mannino, Miro and Le Xuan Anh, Mai and Bagge, Whitney and Abouzied, Azza and Shasha, Dennis},
  booktitle={The 34th Annual ACM Symposium on User Interface Software and Technology},
  pages={894--909},
  year={2021}
}

@article{xu2018ensemblelens,
  title={Ensemblelens: Ensemble-based visual exploration of anomaly detection algorithms with multidimensional data},
  author={Xu, Ke and Xia, Meng and Mu, Xing and Wang, Yun and Cao, Nan},
  journal={IEEE transactions on visualization and computer graphics},
  volume={25},
  number={1},
  pages={109--119},
  year={2018},
  publisher={IEEE}
}

@article{dhanoa2021fuzzy,
  title={Fuzzy spreadsheet: Understanding and exploring uncertainties in tabular calculations},
  author={Dhanoa, Vaishali and Walchshofer, Conny and Hinterreiter, Andreas and Gr{\"o}ller, Eduard and Streit, Marc},
  journal={IEEE Transactions on Visualization and Computer Graphics},
  volume={29},
  number={2},
  pages={1463--1477},
  year={2021},
  publisher={IEEE}
}

@inproceedings{zgraggen2016tableur,
  title={Tableur: Handwritten spreadsheets},
  author={Zgraggen, Emanuel and Zeleznik, Robert and Eichmann, Philipp},
  booktitle={Proceedings of the 2016 CHI Conference Extended Abstracts on Human Factors in Computing Systems},
  pages={2362--2368},
  year={2016}
}

@article{shen2024surroflow,
  title={Surroflow: A flow-based surrogate model for parameter space exploration and uncertainty quantification},
  author={Shen, Jingyi and Duan, Yuhan and Shen, Han-Wei},
  journal={IEEE Transactions on Visualization and Computer Graphics},
  year={2024},
  publisher={IEEE}
}

@inproceedings{bian2021semantic,
  title={Semantic explanation of interactive dimensionality reduction},
  author={Bian, Yali and North, Chris and Krokos, Eric and Joseph, Sarah},
  booktitle={2021 IEEE Visualization Conference (VIS)},
  pages={26--30},
  year={2021},
  organization={IEEE}
}

@article{pajer2016weightlifter,
  title={Weightlifter: Visual weight space exploration for multi-criteria decision making},
  author={Pajer, Stephan and Streit, Marc and Torsney-Weir, Thomas and Spechtenhauser, Florian and M{\"o}ller, Torsten and Piringer, Harald},
  journal={IEEE transactions on visualization and computer graphics},
  volume={23},
  number={1},
  pages={611--620},
  year={2016},
  publisher={IEEE}
}

@inproceedings{splechtna2015interactive,
  title={Interactive visual steering of hierarchical simulation ensembles},
  author={Splechtna, Rainer and Matkovi{\'c}, Kre{\v{s}}imir and Gra{\v{c}}anin, Denis and Jelovi{\'c}, Mario and Hauser, Helwig},
  booktitle={2015 IEEE Conference on Visual Analytics Science and Technology (VAST)},
  pages={89--96},
  year={2015},
  organization={IEEE}
}

@article{cheng2020dece,
  title={Dece: Decision explorer with counterfactual explanations for machine learning models},
  author={Cheng, Furui and Ming, Yao and Qu, Huamin},
  journal={IEEE Transactions on Visualization and Computer Graphics},
  volume={27},
  number={2},
  pages={1438--1447},
  year={2020},
  publisher={IEEE}
}

@article{hao2023timetuner,
  title={TimeTuner: diagnosing time representations for time-series forecasting with counterfactual explanations},
  author={Hao, Jianing and Shi, Qing and Ye, Yilin and Zeng, Wei},
  journal={IEEE Transactions on Visualization and Computer Graphics},
  volume={30},
  number={1},
  pages={1183--1193},
  year={2023},
  publisher={IEEE}
}

@inproceedings{wang2023gam,
  title={Gam coach: Towards interactive and user-centered algorithmic recourse},
  author={Wang, Zijie J and Wortman Vaughan, Jennifer and Caruana, Rich and Chau, Duen Horng},
  booktitle={Proceedings of the 2023 CHI Conference on Human Factors in Computing Systems},
  pages={1--20},
  year={2023}
}

@article{dingen2018regressionexplorer,
  title={RegressionExplorer: Interactive exploration of logistic regression models with subgroup analysis},
  author={Dingen, Dennis and van't Veer, Marcel and Houthuizen, Patrick and Mestrom, Eveline HJ and Korsten, Erik HHM and Bouwman, Arthur RA and Van Wijk, Jarke},
  journal={IEEE transactions on visualization and computer graphics},
  volume={25},
  number={1},
  pages={246--255},
  year={2018},
  publisher={IEEE}
}

@article{zhang2022promotionlens,
  title={Promotionlens: Inspecting promotion strategies of online e-commerce via visual analytics},
  author={Zhang, Chenyang and Wang, Xiyuan and Zhao, Chuyi and Ren, Yijing and Zhang, Tianyu and Peng, Zhenhui and Fan, Xiaomeng and Ma, Xiaojuan and Li, Quan},
  journal={IEEE Transactions on Visualization and Computer Graphics},
  volume={29},
  number={1},
  pages={767--777},
  year={2022},
  publisher={IEEE}
}

@article{wang2024empirical,
  title={An empirical study of counterfactual visualization to support visual causal inference},
  author={Wang, Arran Zeyu and Borland, David and Gotz, David},
  journal={Information Visualization},
  volume={23},
  number={2},
  pages={197--214},
  year={2024},
  publisher={SAGE Publications Sage UK: London, England}
}

@inproceedings{hazarika2023haiva,
  title={HAiVA: Hybrid AI-assisted Visual Analysis Framework to Study the Effects of Cloud Properties on Climate Patterns},
  author={Hazarika, Subhashis and Hirasawa, Haruki and Kim, Sookyung and Ramea, Kalai and Cachay, Salva R and Mitra, Peetak and Hingmire, Dipti and Singh, Hansi and Rasch, Phil J},
  booktitle={2023 IEEE Visualization and Visual Analytics (VIS)},
  pages={226--230},
  year={2023},
  organization={IEEE}
}

@inproceedings{gomez2021advice,
  title={Advice: Aggregated visual counterfactual explanations for machine learning model validation},
  author={Gomez, Oscar and Holter, Steffen and Yuan, Jun and Bertini, Enrico},
  booktitle={2021 IEEE Visualization Conference (VIS)},
  pages={31--35},
  year={2021},
  organization={IEEE}
}

@inproceedings{wang2018seismo,
  title={Seismo: Blood pressure monitoring using built-in smartphone accelerometer and camera},
  author={Wang, Edward Jay and Zhu, Junyi and Jain, Mohit and Lee, Tien-Jui and Saba, Elliot and Nachman, Lama and Patel, Shwetak N},
  booktitle={Proceedings of the 2018 CHI conference on human factors in computing Systems},
  pages={1--9},
  year={2018}
}

@inproceedings{mukashev2023tacttongue,
  title={Tacttongue: Prototyping electrotactile stimulations on the tongue},
  author={Mukashev, Dinmukhammed and Ranasinghe, Nimesha and Nittala, Aditya Shekhar},
  booktitle={Proceedings of the 36th Annual ACM Symposium on User Interface Software and Technology},
  pages={1--14},
  year={2023}
}

@inproceedings{supplementary,
    author = {Anonymous},
    title = {\href{https://osf.io/2ae5v/overview?view_only=141a798ba5b0436bbea0cfa8cb81d151}{Supplementary OSF}},
    author={Anonymous},
}

@inproceedings{hohman2019gamut,
  title={Gamut: A design probe to understand how data scientists understand machine learning models},
  author={Hohman, Fred and Head, Andrew and Caruana, Rich and DeLine, Robert and Drucker, Steven M},
  booktitle={Proceedings of the 2019 CHI conference on human factors in computing systems},
  pages={1--13},
  year={2019}
}

@article{liu2024visualizing,
  title={Visualizing Spatial Semantics of Dimensionally Reduced Text Embeddings},
  author={Liu, Wei and North, Chris and Faust, Rebecca},
  journal={arXiv preprint arXiv:2409.03949},
  year={2024}
}

@inproceedings{browne2020camera,
  title={Camera adversaria},
  author={Browne, Kieran and Swift, Ben and Nurmikko-Fuller, Terhi},
  booktitle={Proceedings of the 2020 CHI Conference on Human Factors in Computing Systems},
  pages={1--9},
  year={2020}
}

@inproceedings{crisan2024exploring,
  title={Exploring Subjective Notions of Explainability through Counterfactual Visualization of Sentiment Analysis},
  author={Crisan, Anamaria and Butters, Nathan and others},
  booktitle={2024 IEEE Evaluation and Beyond-Methodological Approaches for Visualization (BELIV)},
  pages={15--24},
  year={2024},
  organization={IEEE}
}

@article{strobelt2018s,
  title={S eq 2s eq-v is: A visual debugging tool for sequence-to-sequence models},
  author={Strobelt, Hendrik and Gehrmann, Sebastian and Behrisch, Michael and Perer, Adam and Pfister, Hanspeter and Rush, Alexander M},
  journal={IEEE transactions on visualization and computer graphics},
  volume={25},
  number={1},
  pages={353--363},
  year={2018},
  publisher={IEEE}
}

@inproceedings{cavallo2018visual,
  title={A visual interaction framework for dimensionality reduction based data exploration},
  author={Cavallo, Marco and Demiralp, {\c{C}}a{\u{g}}atay},
  booktitle={Proceedings of the 2018 CHI conference on human factors in computing systems},
  pages={1--13},
  year={2018}
}

@article{xie2020visual,
  title={A visual analytics approach for exploratory causal analysis: Exploration, validation, and applications},
  author={Xie, Xiao and Du, Fan and Wu, Yingcai},
  journal={IEEE Transactions on Visualization and Computer Graphics},
  volume={27},
  number={2},
  pages={1448--1458},
  year={2020},
  publisher={IEEE}
}

@inproceedings{booshehrian2012vismon,
  title={Vismon: Facilitating analysis of trade-offs, uncertainty, and sensitivity in fisheries management decision making},
  author={Booshehrian, Maryam and M{\"o}ller, Torsten and Peterman, Randall M and Munzner, Tamara},
  booktitle={Computer Graphics Forum},
  volume={31},
  number={3pt3},
  pages={1235--1244},
  year={2012},
  organization={Wiley Online Library}
}

@article{von2016visualization,
  title={Visualization system requirements for data processing pipeline design and optimization},
  author={Von Landesberger, Tatiana and Fellner, Dieter W and Ruddle, Roy A},
  journal={IEEE Transactions on Visualization and Computer Graphics},
  volume={23},
  number={8},
  pages={2028--2041},
  year={2016},
  publisher={IEEE}
}

@inproceedings{piccolotto2023visual,
  title={Visual parameter space exploration in time and space},
  author={Piccolotto, Nikolaus and B{\"o}gl, Markus and Miksch, Silvia},
  booktitle={Computer Graphics Forum},
  volume={42},
  number={6},
  pages={e14785},
  year={2023},
  organization={Wiley Online Library}
}

@article{espadoto2021unprojection,
  title={UnProjection: Leveraging inverse-projections for visual analytics of high-dimensional data},
  author={Espadoto, Mateus and Appleby, Gabriel and Suh, Ashley and Cashman, Dylan and Li, Mingwei and Scheidegger, Carlos and Anderson, Erik W and Chang, Remco and Telea, Alexandru C},
  journal={IEEE Transactions on Visualization and Computer Graphics},
  volume={29},
  number={2},
  pages={1559--1572},
  year={2021},
  publisher={IEEE}
}

@article{gathani2025if,
  title={What-if Analysis for Business Professionals: Current Practices and Future Opportunities},
  author={Gathani, Sneha and Liu, Zhicheng and Haas, Peter J and Demiralp, {\c{C}}a{\u{g}}atay},
  year={2025}
}

@article{zhao2024leva,
  title={Leva: Using large language models to enhance visual analytics},
  author={Zhao, Yuheng and Zhang, Yixing and Zhang, Yu and Zhao, Xinyi and Wang, Junjie and Shao, Zekai and Turkay, Cagatay and Chen, Siming},
  journal={IEEE transactions on visualization and computer graphics},
  volume={31},
  number={3},
  pages={1830--1847},
  year={2024},
  publisher={IEEE}
}

@inproceedings{setlur2016eviza,
  title={Eviza: A natural language interface for visual analysis},
  author={Setlur, Vidya and Battersby, Sarah E and Tory, Melanie and Gossweiler, Rich and Chang, Angel X},
  booktitle={Proceedings of the 29th annual symposium on user interface software and technology},
  pages={365--377},
  year={2016}
}

@inproceedings{ko2024natural,
  title={Natural language dataset generation framework for visualizations powered by large language models},
  author={Ko, Hyung-Kwon and Jeon, Hyeon and Park, Gwanmo and Kim, Dae Hyun and Kim, Nam Wook and Kim, Juho and Seo, Jinwook},
  booktitle={Proceedings of the 2024 CHI Conference on Human Factors in Computing Systems},
  pages={1--22},
  year={2024}
}

@article{tian2024chartgpt,
  title={Chartgpt: Leveraging llms to generate charts from abstract natural language},
  author={Tian, Yuan and Cui, Weiwei and Deng, Dazhen and Yi, Xinjing and Yang, Yurun and Zhang, Haidong and Wu, Yingcai},
  journal={IEEE Transactions on Visualization and Computer Graphics},
  volume={31},
  number={3},
  pages={1731--1745},
  year={2024},
  publisher={IEEE}
}

@article{luo2021natural,
  title={Natural language to visualization by neural machine translation},
  author={Luo, Yuyu and Tang, Nan and Li, Guoliang and Tang, Jiawei and Chai, Chengliang and Qin, Xuedi},
  journal={IEEE Transactions on Visualization and Computer Graphics},
  volume={28},
  number={1},
  pages={217--226},
  year={2021},
  publisher={IEEE}
}

@inproceedings{tian2024sqlucid,
  title={Sqlucid: Grounding natural language database queries with interactive explanations},
  author={Tian, Yuan and Kummerfeld, Jonathan K and Li, Toby Jia-Jun and Zhang, Tianyi},
  booktitle={Proceedings of the 37th Annual ACM Symposium on User Interface Software and Technology},
  pages={1--20},
  year={2024}
}

@article{fu2023catsql,
  title={Catsql: Towards real world natural language to sql applications},
  author={Fu, Han and Liu, Chang and Wu, Bin and Li, Feifei and Tan, Jian and Sun, Jianling},
  journal={Proceedings of the VLDB Endowment},
  volume={16},
  number={6},
  pages={1534--1547},
  year={2023},
  publisher={VLDB Endowment}
}

@inproceedings{tian2025text,
  title={Text-to-SQL Domain Adaptation via Human-LLM Collaborative Data Annotation},
  author={Tian, Yuan and Lee, Daniel and Wu, Fei and Mai, Tung and Qian, Kun and Sahai, Siddhartha and Zhang, Tianyi and Li, Yunyao},
  booktitle={Proceedings of the 30th International Conference on Intelligent User Interfaces},
  pages={1398--1425},
  year={2025}
}

\end{document}